\documentclass[10pt,prd,amsmath,amssymb,aps,showpacs,twocolumn,unsortedaddress]{revtex4-1}
\newcommand{\be}[1]{\begin{equation}#1\end{equation}}
\newcommand{\ba}[1]{\begin{align}#1\end{align}}
\newcommand{\ban}[1]{\begin{align*}#1\end{align*}}

\newcommand{\lr}[1]{\left( #1\right)}
\newcommand{\exv}[1]{\left< #1\right>}

\newcommand{\dd}{\mathrm{d}}
\renewcommand{\vec}[1]{\mathbf{#1}}

\newcommand{\eqqref}[1]{\text{Eq.}\,\eqref{#1}}
\pdfoutput=1 

\usepackage{hyperref}
\usepackage{graphicx}
\usepackage[normalem]{ulem}
\usepackage[T1]{fontenc}
\usepackage{dcolumn}
\usepackage{bm}
\usepackage[caption=false]{subfig}
\begin{document}

\title{\boldmath Order parameter fluctuations in the holographic superconductor}

\author{N.W.M. Plantz}
\author{H.T.C. Stoof}
\author{S. Vandoren}
\affiliation{Institute for Theoretical Physics and Center for Extreme Matter and Emergent Phenomena, Utrecht University, \\Leuvenlaan 4, 3584 CE Utrecht, The Netherlands}

\email{n.w.m.plantz@uu.nl}
\email{h.t.c.stoof@uu.nl}
\email{s.j.g.vandoren@uu.nl}

\date{\today}

\begin{abstract}
We investigate the effect of order parameter fluctuations in the holographic superconductor. In particular, following an introduction to the concept of intrinsic dynamics and its implementation within holographic models, we compute the intrinsic spectral functions of the order parameter in both the normal and the superconducting phase, using a fully backreacted bulk geometry. We also present a vector-like large-$N$ version of the Ginzburg-Landau model that accurately describes our long-wavelength results in both phases. Our results indicate that the holographic superconductor describes a relativistic multi-component superfluid in the universal regime of the BEC-BCS crossover.
\end{abstract}

\pacs{11.25.Tq, 74.20.De, 74.20.-z}

\maketitle
\flushbottom

\section{Introduction}
\label{sec:intro}
Ginzburg-Landau theory \cite{LG} has been used to describe physics near a conventional superconducting phase transition with great success. Based on the Landau approach to continuous phase transitions, it makes use of a complex order parameter which acquires a nonzero expectation value in the superconducting phase. As this is a phenomenological model, a microscopic interpretation of the order parameter was not included. This interpretation was provided by Gor'kov several years after the Ginzburg-Landau theory \cite{Gorkov}, using the microscopic model of superconductivity by Bardeen, Cooper, and Schrieffer \cite{BCS}. Here, superconductivity is described as the condensation of Cooper pairs, which consist of a pair of electrons on top of a filled Fermi sea, bound together due to a phonon-mediated attractive interaction. By elegantly using a variational \textit{Ansatz} for the BCS ground state, BCS mean-field theory has succeeded in producing many accurate quantitative results that have been confirmed experimentally in weakly coupled superconductors.

As BCS mean-field theory only describes superconductors in the weakly coupled regime, several different approaches have been used to study strongly coupled superconductors. One example of such an approach is Eliashberg theory, which goes beyond the BCS mean-field theory by providing a more accurate treatment of the phonons interacting with the electrons. The self-energy due to these interactions now includes retardation effects, in contrast to the BCS model. Consequently, the Eliashberg formalism is able to provide more accurate quantitative results \cite{Schrieffer} than the BCS formalism. However, in the class of high-temperature superconductors, the pairing mechanism cannot be described by means of interactions with phonons. Therefore even the Eliashberg approach is inapplicable and methods to describe high-temperature superconductors remain mysterious. Fermion gases at unitarity are superconductors at infinite coupling \cite{fn1}. These have been described by numerical approaches based on the quantum Monte Carlo method \cite{QMC1,QMC2,QMC3}. Moreover, an analytical description by means of renormalization group theory can be found in Ref. \cite{RGStoof}.

A novel approach to strongly coupled systems, which has become very popular over the past decade, is the use of the holographic duality. Inspired by ideas in Refs. \cite{Gubser1,Gubser2}, a bottom-up approach of the AdS/CFT correspondence to superconductivity was first given in Ref. \cite{HHH}, followed by many other papers \cite{Intro}. One of the most used models within this framework describes the superconducting phase transition as the condensation of some complex order parameter in the boundary theory that arises from a dual complex scalar field in the classical gravitational theory. From this model, many characteristics of superconductivity have been reproduced, such as the diverging DC conductivity, an energy gap, and a Meissner effect \cite{HHH2}.  The microscopic interpretation of the order parameter is not known, since bottom-up holography usually provides us with expectation values of unknown composite operators rather than the single-particle or single-pair operators which naturally arise in condensed-matter systems. It is therefore unsurprising that results obtained through this approach are generally different from BCS derivations. However, one might wonder whether a phenomenological Ginzburg-Landau theory can still be applied. The answer to this question seems positive, based on e.g. the mean-field critical exponents near the transition temperature \cite{Herzog}.

The long-term aim of this work is to study ultracold fermion gases at unitarity using a holographic approach. The holographic superconductor mentioned above seems like a logical starting point towards this aim, since this model should in principle also consist of strongly correlated fermions. We therefore study the properties of the holographic superconductor in detail in this paper, focussing in particular on the intrinsic dynamics of the order parameter fluctuations in the normal as well as the superconducting phase. While previous studies on the order parameter dynamics mostly concentrate on the poles of the full retarded Green's function \cite{Maeda,Dector}, e.g. by finding the quasinormal modes \cite{Landsteiner}, we directly calculate the intrinsic two-point correlation function of the order parameter, using scalar field fluctuations in the gravity theory. In a mode-coupling theory, the intrinsic dynamics of the order parameter fluctuations corresponds to the dynamics that is uncoupled from the other hydrodynamic degrees of freedom. Notice that this dynamics is different from the full dynamics and can in general not be observed in an experiment. However, the intrinsic dynamics does provide us with important information about the nature of the order parameter. In particular, it enables us to use a gradient expansion and arrive at a local Ginzburg-Landau theory for the holographic superconductor. A similar calculation was performed in Ref. \cite{SchalmZaanen} above the critical temperature. Here we carry out calculations below the critical temperature as well, which include full backreaction in the bulk geometry. Subsequently, we present a modified Ginzburg-Landau model that includes the large-$N$ limit which is implicit in the AdS/CFT duality, and find that our long-wavelength results are well described by this model in both the normal and superconducting phase. However, as a consequence of this large-$N$ limit, we observe that the Higgs mode and the second-sound mode are not present in the intrinsic order parameter fluctuations.

The outline of this paper is as follows. In Section \ref{sec:HSC} , we discuss the background theory that will be used throughout this paper. This includes a short review of the holographic superconductor solution and a comparison with a number of universal BCS results. Moreover, our notation and conventions are introduced here. Section \ref{sec:OPF} starts with a review of the concept of intrinsic dynamics and subsequently covers the scalar field fluctuations on top of the holographic superconductor background, including the resulting intrinsic two-point function of the order parameter in the dual theory. We end with our conclusions in Section \ref{sec:CD} .
\section{The holographic superconductor}
\label{sec:HSC}
In this section, we describe the bulk geometry that we use throughout this paper. This geometry was introduced in Ref. \cite{HHH2}. The purpose of this section is to outline its properties that are most relevant to our results, as well as to introduce our notation and conventions. After giving the bulk solutions, we specify on how many and on which parameters this solution exactly depends. We end the section by discussing the superconducting phase transition that appears in the dual field theory and comparing its properties with universal results which follow from BCS theory.

For the sake of generality, we give the gravitational bulk for an arbitrary spatial dimension $d$. However, we will  always specify to $d=4$ when discussing the dual field theory, which then has three spatial dimensions. Although many high-$T_c$ superconductors consist of layers and are thus effectively two-dimensional, we are interested in three-dimensional superconductors here. Examples of these include the ultracold gases at unitarity mentioned in the introduction.
\subsection{The gravity solutions}
\label{ss:HSCg}
The gravitational background that we use follows from the action that describes gravity minimally coupled to a $U(1)$ gauge field $A_\mu$ and a charged scalar field $\phi$. In SI units, it is given by
\ba{ \label{eq:ActionBR}
S=\int\dd^{d+1}x\sqrt{-g}\bigg[&\frac{c^3}{16\pi G}(R-2\Lambda)
-\frac{1}{4{\mu_0} c}F^2\nonumber \\ &- \lr{|D\phi|^2+\frac{m^2c^2}{\hbar^2}|\phi|^2}\bigg]\,.}
Here the scalar field has mass $m$ and charge $q$. Furthermore, $G$ and $\mu_0$ are Newton's constant and the vacuum permeability in $d$ spatial dimensions respectively. In addition, $\Lambda<0$ is the cosmological constant and $D_{\mu}$ is the gauge covariant derivative
\be{D_{\mu}=\nabla_{\mu}-\frac{iq}{\hbar}A_{\mu}\,.}
The equations of motion that follow from this action describe the gauge field and the scalar field backreacting on the geometry. Here, we consider static solutions to these equations, with planar symmetry. The metric \textit{Ansatz} can then be written as \cite{HHH2}
\be{ \label{eq:metricAnsatz}
\dd s^2=-f(r)e^{-\chi(r)}c^2\dd t^2+\frac{1}{f(r)}\dd r^2+\frac{r^2}{L^2}\dd \vec{x}_{d-1}^2\, ,}
where the AdS radius $L$ is given by $L^2=d(d-1)/ \\ (-2\Lambda)$. Here the coordinate $r$ runs from the horizon $r=r_+$, where $f(r_+)=0$, to the boundary at $r=\infty$. Demanding there to be no conical singularity in the imaginary-time geometry at $r_+$ gives the Hawking temperature
\be{ \label{eq:HawkT}
k_B T = \frac{\hbar c f'(r_+) e^{-\chi(r_+)/2}}{4\pi}\, ,}
where $k_B$ is Boltzmann's constant. Furthermore, the gauge field is temporal, i.e., $A=A_t(r)\dd t$, and we choose a gauge in which $\phi$ is real. With these \textit{Ans\"{a}tze} the equations of motion become
\begin{align}
\phi''+\lr{\frac{f'}{f}+\frac{d-1}{r}-\frac{\chi'}{2}}\phi'-\frac{m^2c^4-q^2A_t^2\frac{e^\chi}{f}}{\hbar^2c^2 f}\phi &=0 \label{eq:KGfT}\, ,\\
A_t''+\lr{\frac{d-1}{r}+\frac{\chi'}{2}}A_t'-2\frac{q^2 \mu_0 c \phi^2}{\hbar^2 f}A_t &=0 \label{eq:MWfT}\, ,\\
\chi'+\frac{32\pi G}{(d-1)c^3}r\lr{\phi'^2+\frac{q^2A_t^2e^\chi}{\hbar^2c^2f^2}\phi^2} &=0 \label{eq:EE1fT}\, ,\\
f'+\lr{\frac{d-2}{r}-\frac{\chi'}{2}}f-\frac{rd}{L^2}\,\,\,\,\quad\qquad &\nonumber \\+\frac{16\pi G}{(d-1)c^3}r\lr{\frac{e^\chi A_t'^2}{2\mu_0 c^3}+\frac{m^2c^2}{\hbar^2}\phi^2}&=0\, . \label{eq:EE2fT}
\end{align}
Our gravitational background consists of solutions to these equations, which we consider next.

Firstly, we consider the solution with a trivial scalar field profile. This is just the well-known Reissner-Nordstr\"{o}m black brane, given by $\phi = \chi = 0$,
\be{ \label{eq:Atnormal}
A_t=\frac{\mu}{q}\bigg[1-\lr{\frac{r_+}{r}}^{d-2}\bigg]}
and
\ba{ \label{eq:fnormal}
f=&\,\frac{r^2}{L^2}-\lr{\frac{r_+}{r}}^{d-2}\frac{r_+^2}{L^2}\nonumber \\&+\frac{8\pi G}{\mu_0 c^6}\frac{d-2}{d-1}\lr{\frac{\mu}{q}}^2\bigg[\lr{\frac{r_+}{r}}^{2(d-2)}-\lr{\frac{r_+}{r}}^{d-2}\bigg].}
This solution exists for any temperature $T$. We have written the solution such that the integration constant $\mu$ has indeed the dimension of energy, consistent with its interpretation as a chemical potential in the dual field theory. A peculiar feature of this solution is that the event horizon and hence the entropy remain nonzero when $T=0$, making this a very unstable phase at low temperatures.

\begin{figure}[!t]
\includegraphics[width=.95\linewidth,clip]{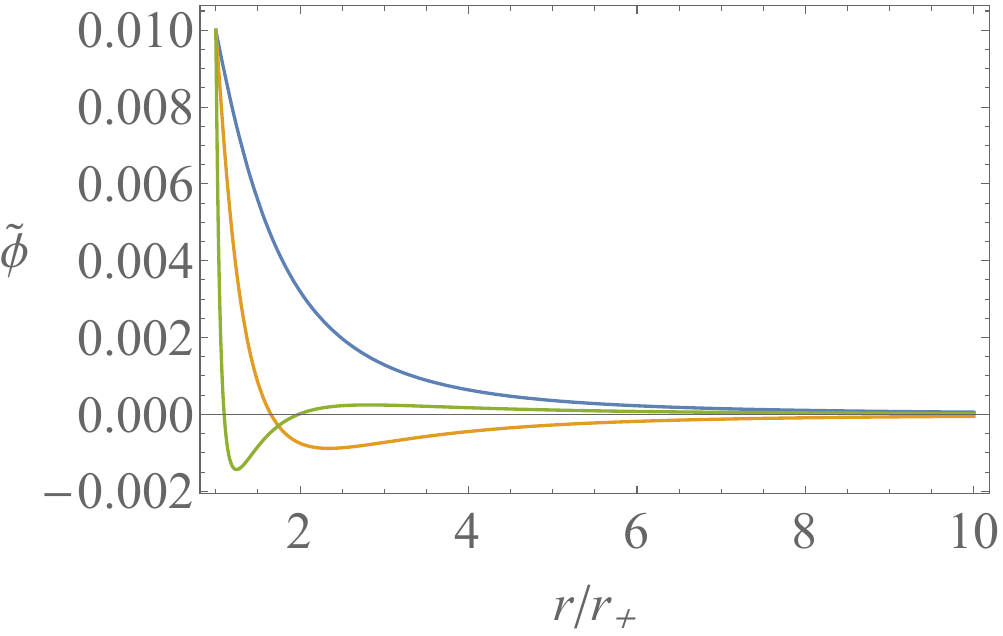}
\caption{\label{fig:phinodes}The scalar field profile for three different solutions of the equations of motion with $\tilde{q}=3$ and $\tilde{m}^2=-3.5$. Here we already use the dimensionless quantities $\tilde{\phi}$, $\tilde{q}$, and $\tilde{m}^2$ defined in \eqqref{eq:dimlessunits}. For each solution the temperature is fixed slightly below the critical temperature, so that $\tilde{\phi}$ remains small. The solution with zero nodes has the highest critical temperature, namely $k_BT_{c0}/\mu\approx 0.075$. The solution with one node has $T_{c1}\approx 0.22 T_{c0}$ and the one with two nodes $T_{c2}\approx0.031 T_{c0}$.}
\end{figure}

The other solution we consider has a nontrivial scalar field. From the equations of motion, we can derive that as $r\rightarrow\infty$, this scalar field behaves as
\be{ \label{eq:phiexp}
\phi=\phi_s \lr{\frac{r}{L}}^{-\Delta_-}+\phi_v \lr{\frac{r}{L}}^{-\Delta_+}+\dots}
with $\Delta_\pm=d/2\pm \nu \equiv d/2 \pm \sqrt{d^2+4\lr{mcL/\hbar}^2}/2$. The particular solutions for which the source $\phi_s=0$ are holographic superconductor solutions in so-called standard quantization \cite{fn2}. Upon imposing the boundary conditions that we discuss in the next subsection, we can numerically find multiple of such solutions which can be characterized by the number of zeros of $\phi$. These hairy black branes only exist below a certain critical temperature $T_c$, which is proportional to $\mu$ and depends on $d$, $m^2$, and $q$. Keeping $\mu$ fixed, we have checked that the solutions where the scalar field has no nodes have the highest critical temperature. In Fig. \ref{fig:phinodes} three solutions for $\phi$ with a different number of nodes are plotted just below their critical temperature.

Note that the parameter $m^2$ is constrained by the Breitenlohner-Freedman bound to $\lr{mcL/\hbar}^2>-d^2/4$, so that the coefficients $\Delta_{\pm}$ are always real. Moreover, we will restrict ourselves to $\lr{mcL/\hbar}^2\leq-d^2/4+1$. By doing so, we can compare our results in the following section to results where alternative quantization is used. Within this range for $m^2$, hairy black brane solutions should exist for any $q$ \cite{ZeroT}. However, finding solutions for $q<1$ turned out to be very difficult numerically.

Choosing the solutions with the lowest thermodynamic potential for a fixed $\mu$, i.e., using the grand-canonical ensemble, our gravitational background is given by the Reissner-Nordstr\"{o}m solution for temperatures above $T_c$. Below $T_c$, the hairy black brane where the scalar field has no zeros is thermodynamically favorable \cite{Herzog}. We can show that for $T=0$, the event horizon of this hairy black brane vanishes \cite{ZeroT}, so that we no longer suffer from the abovementioned instability of the Reissner-Nordstr\"{o}m solution.
\subsection{Free parameters and boundary conditions} \label{ss:param}
An important property of the solutions is the number of parameters we can tune. Hence we proceed by listing the boundary conditions imposed on the solutions. First of all, we introduce the following dimensionless fields and coordinates:
\be{ \label{eq:dimlessunits}
\begin{cases}
(\tilde{t},\tilde{\vec{x}},\tilde{r})=(ct,\vec{x},r)/L \\
\tilde{m}=\frac{cL}{\hbar}m \\
\tilde{A}_{\tilde{t}}=\sqrt{\frac{16\pi G}{\mu_0 c^6}}A_t \\
\tilde{\phi}=\sqrt{\frac{16\pi G}{c^3}}\phi \\
\tilde{q}=\sqrt{\frac{\mu_0 c^6 }{16\pi G}}\frac{L}{\hbar c}q. \\
\end{cases}}
Notice that this eliminates $G$, $\mu_0$, and $L$ from the equations of motion. In the remainder of this paper we will only use dimensionless units derived from the ones above, while omitting the tildes on the quantities. This implies that energy scales, such as $qA_t$, $\mu$, and $k_B T$, are measured in units of $\hbar c/L$, whereas all length scales are measured in units of $L$. The results can easily be converted back to SI units using \eqqref{eq:dimlessunits}.

Upon introducing the dimensionless quantities from \eqqref{eq:dimlessunits} in the action in \eqqref{eq:ActionBR}, we obtain that
\be{ \label{eq:Ndef}
S/\hbar=\frac{c^3 L^{d-1}}{16\pi G\hbar}\tilde{S}\equiv N_G \tilde{S},}
where $\tilde{S}$ is the dimensionless action that no longer explicitly contains the quantities $G$, $\mu_0$, and $L$. Hence, the action is proportional to the dimensionless constant $N_G$, which is related to the integer $N$ of the large-$N$ limit of the dual field theory. The dimensionless quantities in \eqqref{eq:dimlessunits} are defined exactly such that they do not depend on $N$. However, some SI quantities in the action necessarily contain a dependence on $N$, and therefore on $G$, as we will see later on.

Given a dimension $d$, the remaining parameters that determine the bulk geometry are $m^2$ and $q$. Furthermore, as the equations of motion are of first order for $\chi$ and $f$ and of second order for $A_t$ and $\phi$, we need six initial conditions for a particular solution. Finally, we have the position of the event horizon $r_+$, at which we will impose the initial conditions.

Two conditions at the event horizon are given by $A_t(r_+)=0$ and $f(r_+)=0$. Furthermore, multiplying \eqqref{eq:KGfT} by $f$ and evaluating at $r_+$ yields the constraint
\be{
f'(r_+)\phi'(r_+)=m^2\phi(r_+),}
leaving three initial conditions. Requiring the solution to be asymptotically AdS implies requiring that $\chi(\infty)=0$. This condition can be incorporated by first using the initial condition $\chi(r_+)=0$ and afterwards rescaling the solution using the symmetry
\be{ \label{eq:scaleC}
e^\chi\rightarrow C^2 e^\chi, \qquad t \rightarrow Ct, \qquad A_t \rightarrow A_t/C,}
with $C=e^{-\chi(\infty)/2}$, which leaves the equations of motion invariant. Finally, we fix another initial condition by requiring $\phi_s=0$ in \eqqref{eq:phiexp}, corresponding to an unsourced vacuum expectation value. We are thus left with only one initial condition that is unspecified.

Using another symmetry of the equations of motion given by
\be{
\label{eq:scalerp}
r\rightarrow ar, \quad (t,\vec{x}) \rightarrow (t,\vec{x})/a, \quad f \rightarrow a^2 f, \quad A_t \rightarrow a A_t,}
we can obtain any solution from a solution with $r_+=1$. Thus we see that our bulk solution can only nontrivially depend on $d$, $m^2$, $q$ and, due to the unspecified initial condition, on one additional parameter which we take to be $k_BT/\mu$.

Naturally, the geometry of the Reissner-Nordstr\"{o}m black brane does not depend on the parameters $m^2$ and $q$. The dependence on these parameters becomes visible only after including scalar fluctuations to this background.
\subsection{The phase transition and Ginzburg-Landau theory} \label{ss:pt}
To describe the theory on the boundary, we concentrate on the case $d=4$, such that the boundary has three spatial dimensions. After having solved the equations of motion, we can extract boundary values corresponding to physical quantities from the solution. From the scalar field expansion \eqqref{eq:phiexp}, we obtain the order parameter $\left<O\right>=2\nu\phi_v$ which is sourced by $\phi_s$, see e.g. Ref. \cite{Hartnoll}. Similarly, we can expand the gauge field near the boundary as
\be{ \label{eq:Atexp}
A_t = \frac{\mu}{q}- \frac{n q}{2} r^{-2} +\dots\, .}
Here we have written the integration constants in such a way that $\mu$ corresponds to the dimensionless chemical potential, measured in units of $\hbar c/L$. The quantity $n$ corresponds to a dimensionless number density in the dual field theory. Finally, given a bulk solution, we obtain the temperature of the dual field theory from the Hawking temperature in \eqqref{eq:HawkT}.

From the discussion in the previous subsection it follows that for a given $q$ and $m^2$ physical quantities can depend on $k_BT/\mu$, but may also contain a dependence on $N_G$. This dependence follows directly from the proportionality of the bulk action to $N_G$ in \eqqref{eq:Ndef}. Additionally, we may wonder about the physical meaning of the bulk parameters. The mass $m$ determines the scaling dimension of the order parameter $\exv{O}$, as follows from the expansion in \eqqref{eq:phiexp}. The charge $q$ also defines a property of the field theory. In particular, it is related to the structure constants that appear in the three-point functions \cite{Gubser2}. Note that the dimensionless charge $q$ does not give the charge of the operator $\exv{O}$, which can most easily be seen from its definition in \eqqref{eq:dimlessunits}. As we have no numerical value of the AdS radius $L$ and the constants $\mu_0$ and $G$ in $d=4$ dimensions, the proportionality factor between the charge in SI units and its dimensionless counterpart remains unknown. Thus, even if we consider $\exv{O}$ as an expectation value related to Cooper pairs, so that we know there are two particles involved, we do not know the value of $q$. Finally, we have the parameter $N_G$, which is proportional to the integer $N$ of the large-$N$ limit. Hence in this limit we still have a finite parameter $N/N_G$.

\begin{figure}[!t]
\subfloat[$q=3$\label{fig:OTmuq3}]
	{\includegraphics[trim = 0.5cm 0cm 0cm 0cm, clip=true, scale=.25]{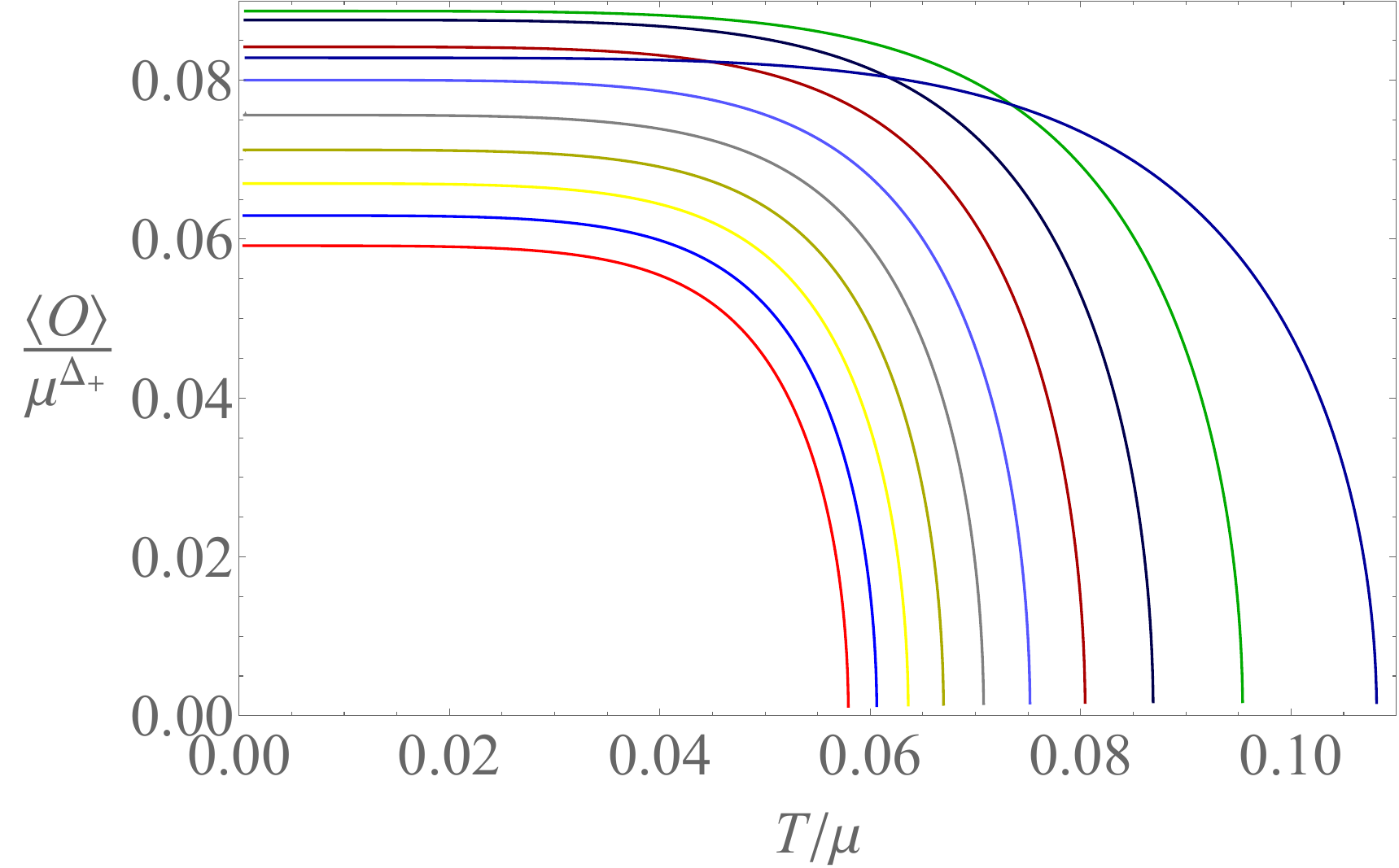}}

\subfloat[$m^2=-3.5$]
	{\includegraphics[trim = 0cm 0cm 0cm 0cm, clip=true, scale=.4]{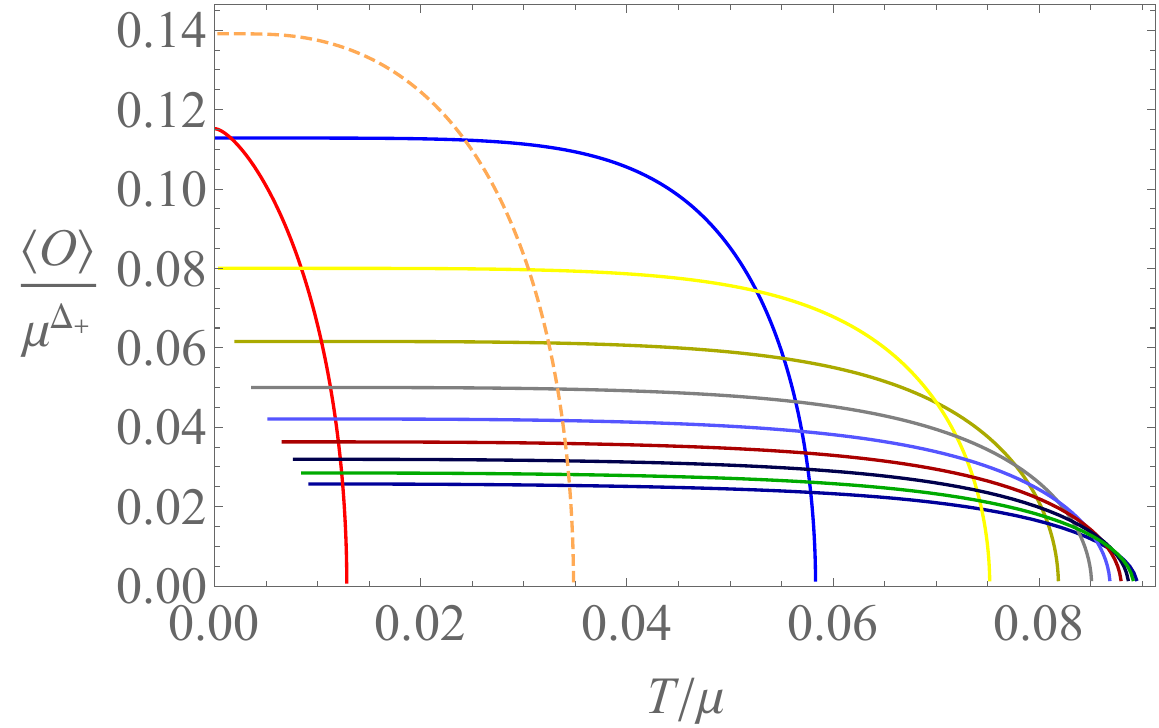}}
\caption{(color online) The expectation value of the order parameter as a function of the temperature. In (a), $q$ is fixed and $m^2$ decreases from $m^2=-3$ for the curve with the lowest critical value of $T/\mu$ to $m^2=-3.9$ for the curve with the highest critical value of $T/\mu$, with steps $\Delta m^2=-0.1$. In (b), $m^2$ is fixed and $q$ increases from $q=1$ for the curve with the lowest critical value of $T/\mu$ to $q=10$ for the curve with the highest critical value of $T/\mu$. We plotted integer values of $q$ here. The exception is the dashed orange curve for $q=1.4$, which was added to show that the dependence on $q$ of the expectation value of the order parameter at zero temperature is not monotonic.}
\label{fig:OTmu}
\end{figure}

As the source term in the background is put to zero by construction, the dual theory acquires an unsourced expectation value below $T_c$. This corresponds to an order parameter of a phase transition which spontaneously breaks the $U(1)$ symmetry. In Fig. \ref{fig:OTmu} the order parameter is shown as a function of the ratio $k_B T/\mu$ for various values of $q$ and $m^2$. We can deduce from this that the phase transition is always of second order. Therefore, we can describe the order parameter $\left<O\right>$ with a Ginzburg-Landau model. More specifically, such a model can be represented by the action 
\be{ \label{eq:LGaction}
S=-\int\dd t\int\dd^{3} \vec{x} \lr{\alpha|O|^2+\frac{\beta}{2}|O|^4},}
where $\alpha$ and $\beta$ are temperature-dependent real coefficients. For $\beta>0$, there appears a nontrivial global minimum
\be{ \label{eq:expv}
\left<O\right>=\sqrt{-\frac{\alpha}{\beta}},}
when $\alpha$ becomes negative below the transition temperature. Note that we have chosen the expectation value of $O$ to be real here, which corresponds to the gauge choice of a real $\phi$ in the bulk theory. Moreover, the numerical data yields that $\left<O\right>\propto |T-T_c|^{1/2}$ near $T_c$. This suggests the conventional choice $\alpha(T)\approx \alpha_0 (T-T_c)$ and $\beta=\beta_0\neq 0$ for the coefficients near $T_c$, as from this we indeed obtain the well-known mean-field result $\left<O\right>\propto |T-T_c|^{1/2}$ in the superconducting phase. The temperature dependence of $\alpha$ and $\beta$ is confirmed by the calculations performed below. However, it is in this bottom-up case not possible to extract such a temperature dependence from a microscopic theory, since as mentioned in the introduction, the microscopic origin of the order parameter is not known. It may however be possible to formulate a microscopic theory using a top-down approach, where one starts with the full duality between type IIB string theory and super Yang-Mills theory, and performs consistent truncations to arrive at the desired model. 

\begin{figure}[!t]
\subfloat[Critical temperature\label{fig:Tc}]
	{\includegraphics[trim = 0.5cm 0.1cm 0cm 0cm, clip=true, scale=.27]{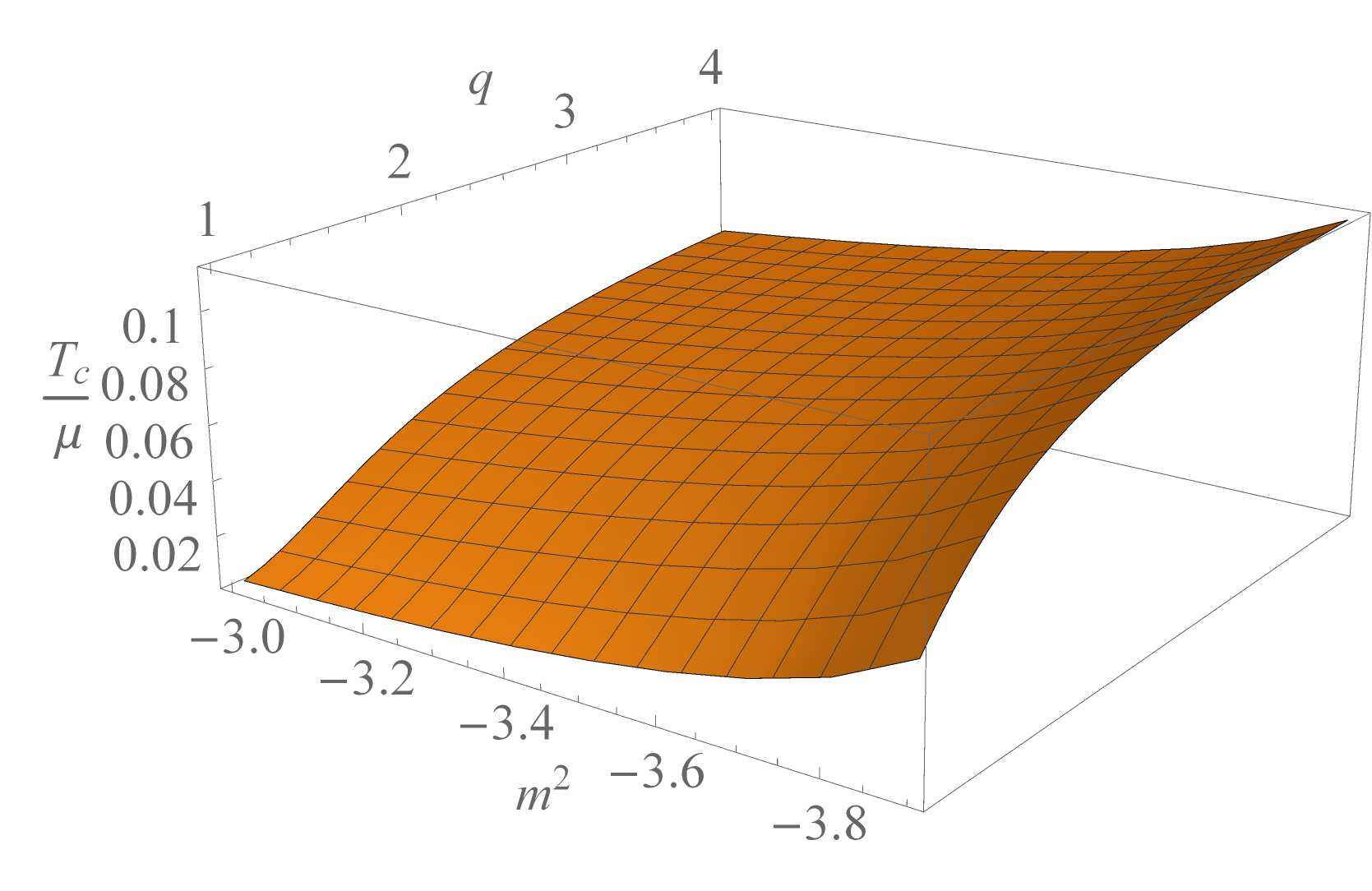}}

\subfloat[$\alpha_0/\beta_0$\label{fig:propTc}]
	{\includegraphics[trim = 0cm 0.1cm 0cm 0cm, clip=true, scale=.27]{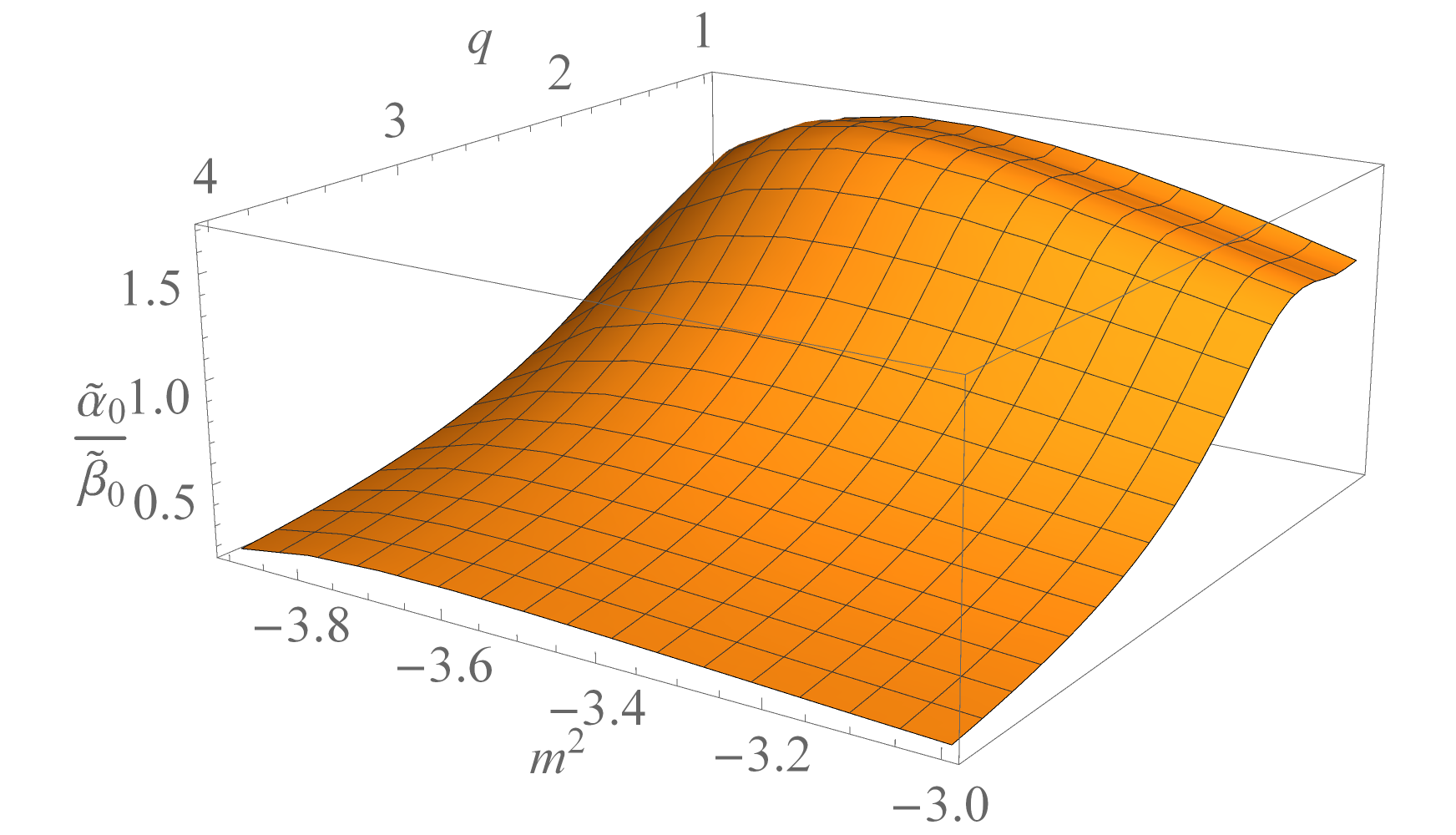}}
\caption{(a) The critical temperature as a function of the parameters $q$ and $m^2$. (b) The proportionality constant between the order parameter and $|T-T_c|^{1/2}$ near the critical temperature as a function of $q$ and $m^2$.  The tildes above the parameters imply that they are scaled with appropriate powers of $\mu$ to make their scaling dimension zero.}
\label{fig:Tcandprop}
\end{figure}

\begin{figure}[!t]
\includegraphics[width=.95\linewidth,clip]{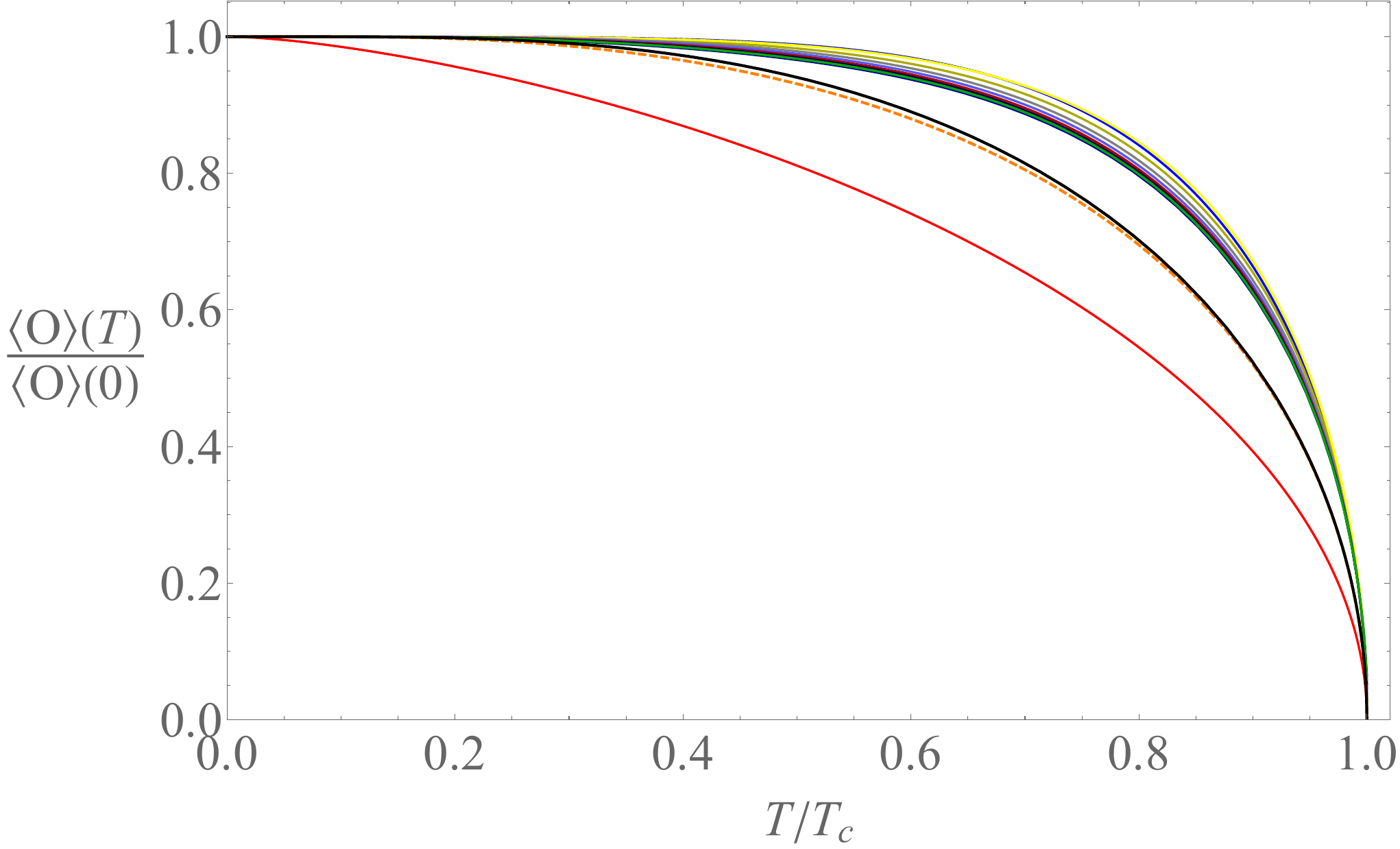}
\caption{(color online) The scaled order parameter $\exv{O}/\exv{O}_{T=0}$ as a function of $T/T_c$ for different values of $q$ and $m^2$. We have used all values of $q$ and $m^2$ which are plotted in Fig. \ref{fig:OTmu} as well. The black curve is the result from BCS theory. Apart from the red curve and dashed orange curve, corresponding to $q=1$ and $q=1.4$ respectively and $m^2=-3.5$, the dependence on $q$ and $m^2$ seems small. The other curves have $q\geq 2$.}
\label{fig:scaled}
\end{figure}

Although the above Ginzburg-Landau model can be applied for all the cases shown in Fig. \ref{fig:OTmu} , the parameters in the model can be seen to depend on $q$ and $m^2$. For example, the value of the order parameter at zero temperature has a nontrivial dependence on both $m^2$ and $q$. The figure shows that this dependence is not monotonic. In general, the critical temperature increases upon increasing $q$ or $|m^2|$, as shown in Fig. \ref{fig:Tc} \cite{fn3}. The coefficients $\alpha$ and $\beta$ also depend on $q$ and $m^2$. In Fig. \ref{fig:propTc} we have shown this dependence for the coefficient $\alpha_0/\beta_0$, which is the square of the proportionality constant between $\exv{O}$ and $|T-T_c|^{1/2}$ near the transition temperature. Scaling the temperature by $T_c$ and the order parameter by its value at $T=0$, we obtain from Fig. \ref{fig:OTmu} the plots in Fig. \ref{fig:scaled} . We see that the rescaled curves show very little dependence on $m^2$ and $q$, except for lower values of $q$. The black curve corresponds to BCS theory. This is a universal result, i.e., this curve is common to all BCS superconductors. Hence, deviations from this curve are a result of strong-coupling effects.

In BCS theory, the order parameter corresponds to the energy gap $\Delta$ of the fermionic single-particle excitation spectrum. The quantity $\left<O\right>$ does not in general have the (scaling) dimension of energy. Nevertheless, since both $\Delta$ and $\exv{O}$ show mean-field behavior near the transition temperature, $\exv{O}$ could be proportional to the gap. The proportionality constant can still depend on $q$ and $m^2$ in a nontrivial way. This proportionality constant should cancel in Fig. \ref{fig:scaled} , i.e., $\exv{O}/\exv{O}_{T=0}=\exv{\Delta}/\exv{\Delta}_{T=0}$. We have therefore attempted to extract this proportionality constant from this figure in a different manner, using the fact that at small temperatures we have the behavior \cite{Kleinert}
\be{ \frac{\exv{\Delta}(T)}{\exv{\Delta}(0)}-1\propto \exp\big[-\exv{\Delta}(0)/T\big].}
However, as we approached zero temperature, our numerical data became too inaccurate to reliably obtain the gap from this expression.
\begin{figure}[!t]
\subfloat[The number density at zero temperature\label{fig:n0}]
	{\includegraphics[trim = 0.2cm 0.1cm 0cm 0cm, clip=true, scale=.25]{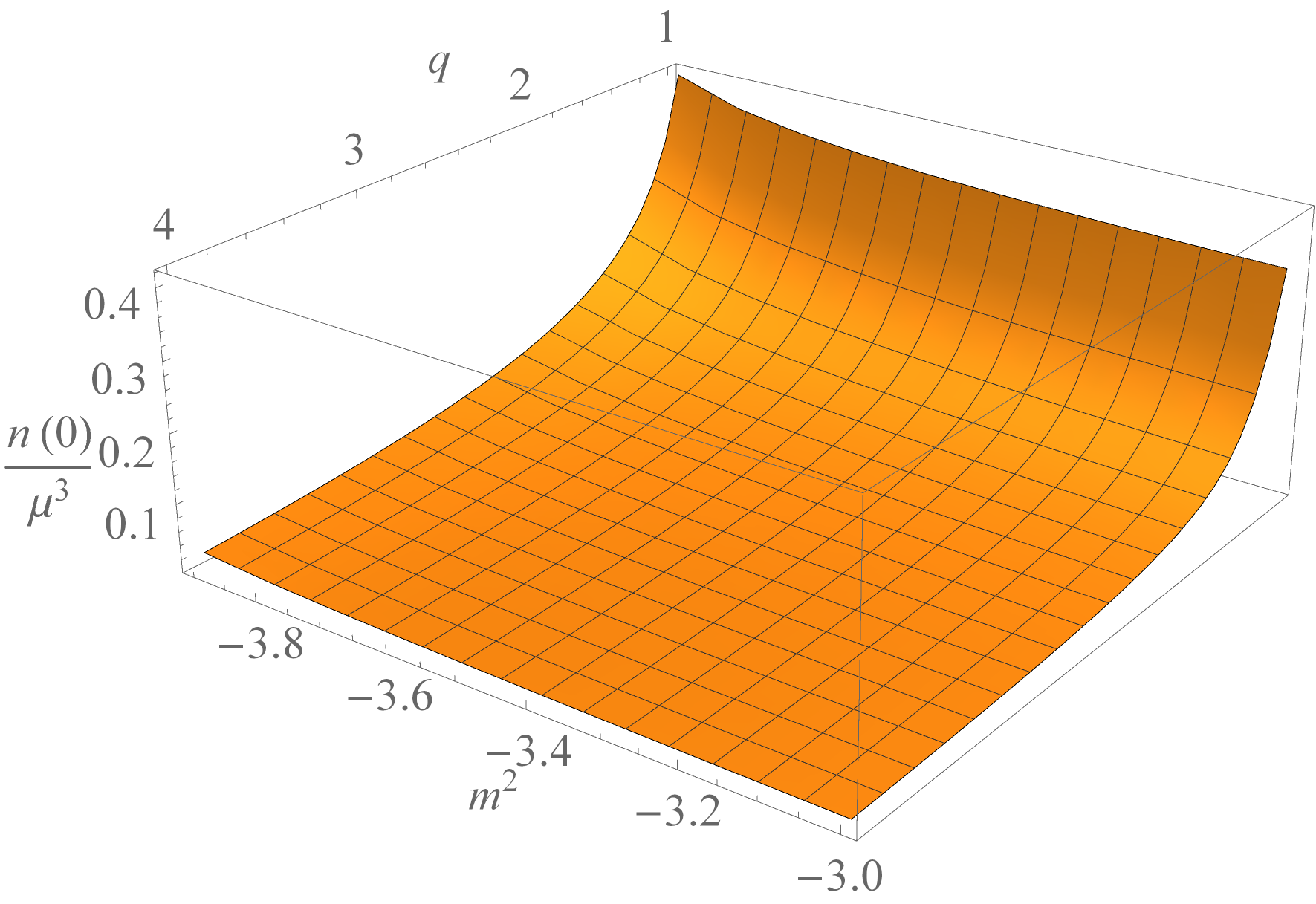}}

\subfloat[The parameter $\beta$\label{fig:beta}]
	{\includegraphics[trim = 1cm 0.1cm 0cm 0cm, clip=true, scale=.26]{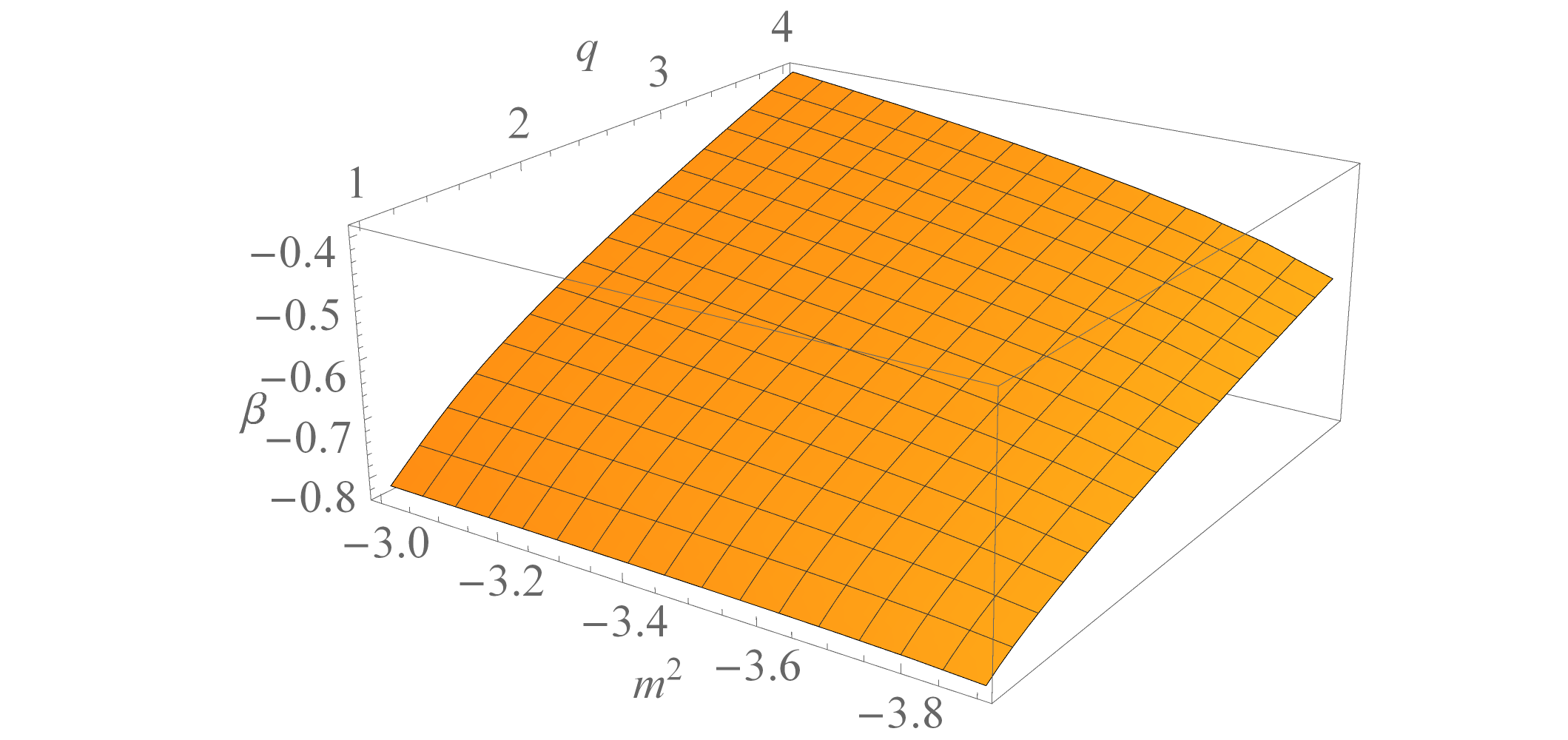}}
\caption{(a) The total number density of the holographic superconductor at zero temperature. (b) The parameter $\beta\equiv-1+\mu/\epsilon_F$ as a function of $q$ and $m^2$. Here we have taken $N/N_G=1$.}
\label{fig:betadata}
\end{figure}

Finally, we have shown in Fig. \ref{fig:n0} the dependence of the zero-temperature (dimensionless) number density $n(0)$ on the parameters $q$ and $m^2$. This number density is determined from the bulk solution using \eqqref{eq:Atexp}. Notice that for a given $q$ and $m^2$, the ratio $n(0)/\mu^3$ is fixed. In contrast, since the action is proportional to $N_G$ as defined in \eqqref{eq:Ndef}, the total density in SI units contains an additional factor of $N_G$ with respect to its dimensionless counterpart. The exact relation is
\be{ \label{eq:weirdn}
n=N_G\tilde{n}L^{-3},}
where we temporarily restored the tilde to distinguish the dimensionless density $\tilde{n}$ from the dimensionful one $n$. It follows that the total density diverges in the large-$N$ limit. However, we are interested in the density of one species, i.e., the total density divided by the number of species $N$. This density coincides with the density numerically obtained from \eqqref{eq:Atexp} up to the factor $N_G/N$, which remains an unknown parameter, but should in principle be fixed by a top-down approach.

The fixed value of $n(0)/\mu^3$ is reminiscent of an ultracold fermion gas near a Feshbach resonance \cite{Tiesinga}. In such an ultracold gas, there are two independent length scales at zero temperature. One of these is the \textit{s}-wave scattering length $a$, which controls the strength of the interaction between the fermions within a Cooper pair. The other length scale is the inverse Fermi wavelength $k_F^{-1}$, which at zero temperature is related to the number density by $n=k_F^3/(3\pi^2)$ for a single fermion species with two spin components. All dimensionless thermodynamic quantities can then be expressed as a function of the dimensionless quantity $1/k_Fa$.  In the weakly coupled BCS limit there are small attractive interactions, so that $1/k_Fa$ becomes very negative, whereas in the BEC limit $1/k_Fa$ is positive and the Cooper pairs form two-body bound states. In the strongly coupled regime $1/k_F|a|<1$ there is a smooth crossover between the BEC and BCS regime, which is appropriately called the BEC-BCS crossover. In the unitarity limit, $1/k_Fa=0$ as $a$ diverges, such that the thermodynamics can only depend on $k_F$. Similar to the situation in our dual field theory, dimensionless quantities like $\mu/\epsilon_F\equiv 1+\beta$, with $\epsilon_F$ the Fermi energy, then become universal constants. This is one of the claims of the so-called universality hypothesis \cite{Ho}.

The function $\beta(k_F a)$, not to be confused with the parameter $\beta$ in the Ginzburg-Landau action in \eqqref{eq:LGaction}, can be determined from experiments. For an ideal gas at zero temperature one has $\mu=\epsilon_F$, so that $\beta=0$. In the weakly coupled BCS limit one has small attractive interactions, so that $\beta$ becomes a small negative number. For an ultracold Fermi gas at unitarity, the variational BCS wave function yields that $\beta=-0.4$ \cite{UQF}, whereas Monte-Carlo simulations show that $\beta\approx -0.6$ \cite{QMC1} and experiments have yielded $\beta=-0.7 \pm 0.1$ \cite{Bartenstein}. Naturally the strong coupling yields a deviation from the BCS theory result. In Fig. \ref{fig:beta} , we have plotted the constant $\beta$ for the holographic superconductor. To obtain this figure, we assume that the bosonic order parameter comes from a pair of fermions \cite{fn4} with Fermi velocity $c$, such that $\epsilon_F=\hbar c k_F$. This is because $\epsilon_F$ is defined with respect to the reference system dual to AdS spacetime without hair, which yields a relativistic field theory where the Dirac cones are just given by $\omega=\pm c|\vec{k}|$. Moreover, we have fixed the number of species to $N=N_G$. For the values of $q$ and $m^2$ shown, we see that the result obtained in Fig. \ref{fig:beta} has the right order of magnitude for a superfluid in the BEC-BCS crossover regime.

\begin{figure}[!t]
\subfloat[$T_c/\mu$\label{fig:Tcsat}]
	{\includegraphics[trim = 0.2cm 0cm 0cm 0cm, clip=true, scale=.4]{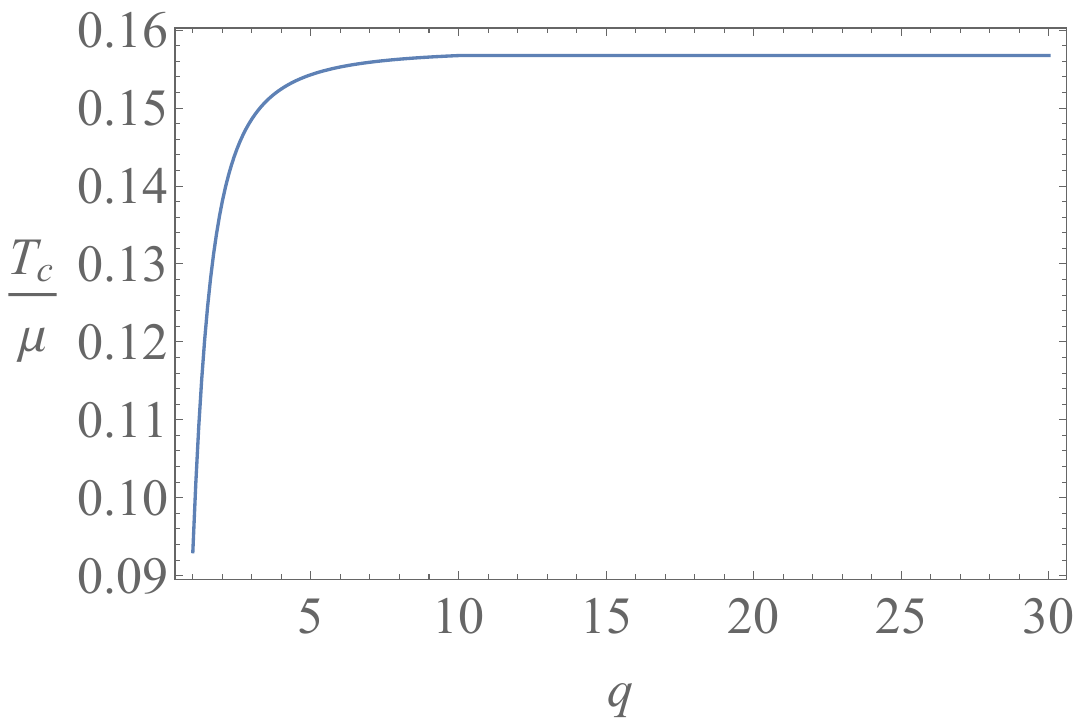}}

\subfloat[$T_c/\epsilon_F$\label{fig:Tcsatef}]
	{\includegraphics[trim = 0cm 0cm 0cm 0cm, clip=true, scale=.4]{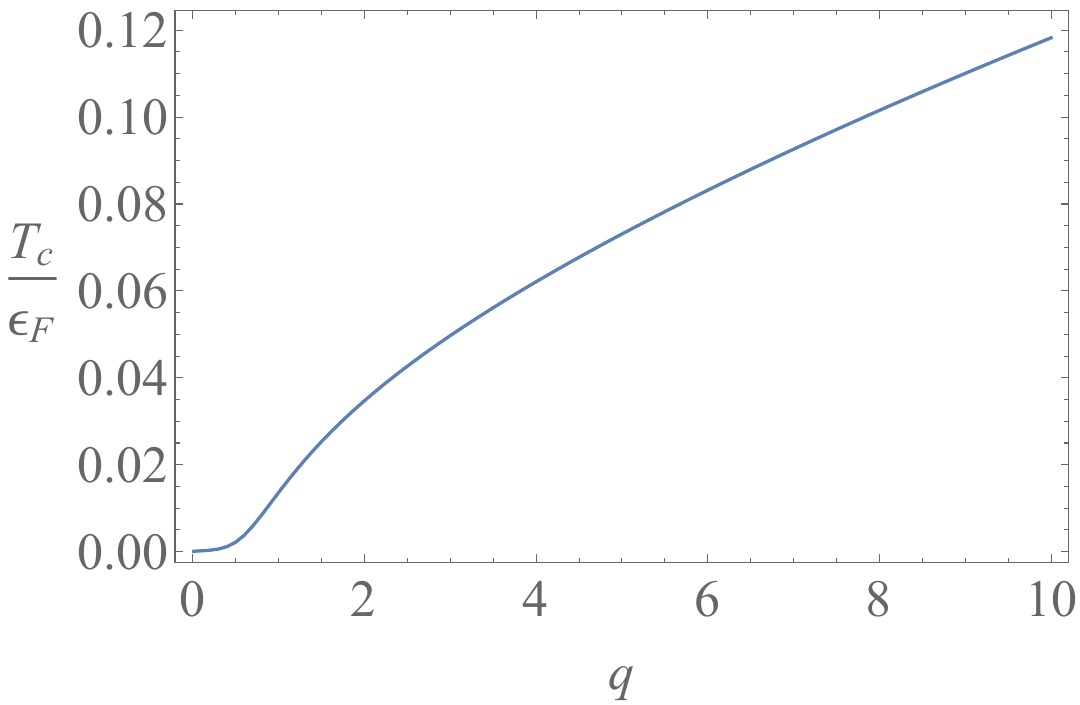}}
\caption{The critical temperature scaled by (a) the chemical potential and (b) the Fermi energy respectively. In both figures, $m^2=-3.9999$. In (b), we used $N/N_G=1$.}
\label{fig:Tchigh}
\end{figure}

The question now arises how high the critical temperature of the strongly coupled superconductor in the dual field theory actually is. As we have already noticed from Fig. \ref{fig:Tc} , the critical temperature is highest when $m^2$ is close to the BF-bound and $q$ is large. As shown in Fig. \ref{fig:Tcsat} , the critical ratio of $T$ and $\mu$ saturates for large $q$ to $T_c/\mu\approx 0.16\approx 1/2\pi$. These values are comparable to the regime of unitary Fermi gases. If we rescale $T_c$ by the Fermi energy instead, using the same value for $N_G/N$ as before, we obtain Fig. \ref{fig:Tcsatef} . We obtain that $T_c/\epsilon_F\propto q^{2/3}$ for large values of $q$. Due to this dependence on $q$ and $N_G/N$, we cannot unambiguously compare the value of $T_c/\epsilon_F$ to other results, obtained for example by experiments or quantum Monte-Carlo simulations. Finally we note that in alternative quantization, the critical temperature is the highest when $m^2$ is close to the upper bound $m^2=-3$. Then the ratio $T_c/\mu$ saturates to about $1.7$, which is more than ten times larger than in normal quantization.
\section{Order parameter fluctuations} \label{sec:OPF}
In the previous section, we have specified which gravitational background and corresponding dual field theory we use. In this section, we study the scalar field fluctuations on top of this background. From these we subsequently determine the two-point function corresponding to the intrinsic order parameter dynamics of the dual field theory and try to describe this with a time-dependent Ginzburg-Landau model. Before doing so, we will start by making clear what we exactly mean by the \textit{intrinsic} dynamics and in what way it differs from the \textit{full} dynamics.
\subsection{Full dynamics} \label{ss:FD}
When given an action for the bulk theory, there is a well known procedure for calculating the corresponding retarded Green's function of the boundary. This gives information about the dynamics of the system, e.g., the quasinormal modes coincide with the poles of the retarded Green's function \cite{SonStarinets}. We will refer to the dynamics found by this procedure as the \textit{full} dynamics. Moreover, as we will explain in the next subsection, this does not coincide with the \textit{intrinsic} dynamics which we consider in this paper.

Let us first sketch how to find the retarded Green's function with the full dynamics of the system. A more extensive explanation of this approach can be found e.g. in Ref. \cite{Kaminski}. Consider a bulk action that depends on the fields $\Phi^I$. Here, the index $I$ labels the different fields in the theory, which are in our case the scalar field $\phi$, the gauge field $A_\mu$, and the metric $g_{\mu\nu}$. To obtain the Green's function, we expand the action up to second order in the fluctuations $\delta\Phi^I$ of the bulk fields around their expectation values $\exv{\Phi^I}$. The result can be written in the form
\be{ \label{eq:S2}
S^{(2)}=-\frac{1}{2}\int \dd^{d+1}x \,\delta\Phi^\dagger \textbf{G}^{-1}_B\delta\Phi+S^{(2)}_\partial,}
where $S^{(2)}_\partial$ is a boundary action and $\textbf{G}^{-1}_B$ is a linear operator acting on $\delta\Phi$. This defines the linearized equations of motion in the bulk as 
\be{ \label{eq:leom}
\textbf{G}^{-1}_B\delta\Phi=0.}
We refer to the Appendix for an explicit example. Because the matrix $\textbf{G}^{-1}_B$ is in general not diagonal, this becomes a coupled system of linear ordinary differential equations.

Near the boundary, we can expand the solutions to the linearized equations of motion as 
\be{ \label{eq:Phiexp}
\delta\Phi^I=\delta\Phi_s^I r^{-\Delta_-^I}+\delta\Phi_v^I r^{-\Delta_+^I}+\dots\,.}
Here, the values of the exponents $\Delta_{\pm}^I$ depend on which field is considered. For example, for the scalar field the exponents are given in \eqqref{eq:phiexp}. Furthermore, the coefficients $\delta\Phi^I_v$ correspond to fluctuations in the expectation values of the operators in the dual field theory. The $\delta\Phi_s^I$ are the source fluctuations. In general, the fluctuation of a single source will lead to fluctuations in the expectation value of all operators. This is a direct consequence of the fact that the linearized equations of motion in \eqqref{eq:leom} are coupled. Writing the fluctuations in Fourier space as 
\be{
\delta\Phi^I(r,\vec{x},t)=\int\frac{\dd^d k}{(2\pi)^d}\delta\Phi^I(r,k) e^{ik_ax^a},}
with the dimensionless four-momentum $k_a=(-\omega,\vec{k})$ \cite{fn5}, we can write the boundary action $S^{(2)}_\partial$ in the form
\be{\label{eq:S2bdy}
S^{(2)}_\partial=\frac{1}{2}\frac{1}{(2\pi)^d}\int\dd\omega\int \dd^{d-1} \vec{k}\, \delta\Phi_s^\dagger \textbf{G}_R\delta\Phi_s.}
The matrix $\textbf{G}_R(\omega,\vec{k})$ is the retarded Green's function of the boundary theory in Fourier space. It describes the dependencies of the fluctuations of the expectation values $\phi_v$ on the fluctuations of the sources $\phi_s$. These are in turn found from the solutions to the linearized equations of motion in \eqqref{eq:leom}, using infalling boundary conditions at the horizon in order to get the \textit{retarded} Green's function \cite{SonStarinets}. 

Now, suppose that we wish to study the full dynamics of the order parameter fluctuations by computing the correlator $\exv{{O'}^*O'}$, which is proportional to one of the components of the matrix $\textbf{G}_R$ in \eqqref{eq:S2bdy}. By computing the on-shell action, it is shown in the Appendix that
\be{
S^{(2)}_\partial=\frac{1}{2}\frac{1}{(2\pi)^d}\int\dd\omega\int\dd^{d-1} \vec{k} [2\nu\lr{\delta\phi_s^*\delta\phi_v+\text{h.c.}}]+\dots\,,}
where we remind the reader that (see after \eqqref{eq:phiexp}) $\nu\equiv\sqrt{d^2+4\lr{mcL/\hbar}^2}/2$. The bulk action does not contribute, as it vanishes due to the linearized equations of motion in \eqqref{eq:leom}. The terms represented by the dots are associated with contributions from the other fields, but cannot yield any terms proportional to $\delta\phi_s^*\delta\phi_s$. Moreover, the hermitian conjugate (h.c.) terms contribute to the correlator $\exv{O'{O'}^*}$. From the expression above it follows that 
\be{ \label{eq:varphi}
i\exv{{O'}^*O'}=2\nu\lr{\frac{\partial\delta\phi_v}{\partial\delta\phi_s}}_{\delta\Phi_s^{I\neq\phi}}.}
In the Appendix we demonstrate how to derive this in the case of the probe limit. This expression denotes the variation of $\delta\phi_v$ with respect to $\delta\phi_s$, where the other sources are kept constant under this variation. We can calculate this as follows. Since in general the order parameter fluctuations are influenced by variations of all the sources in the theory, we can write the scalar field fluctuations near the boundary as
\ba{ 
\delta\phi&=\delta\phi_s  r^{-\Delta_-}+\delta\phi_v r^{-\Delta_+}+\dots \nonumber\\
&= \delta\phi_s  r^{-\Delta_-}+a_I\delta\Phi_s^I r^{-\Delta_+}+\dots\,\label{eq:deltaphiexp},}
where the sum over $I$ is over all the field fluctuations and where $a_I$ are frequency and momentum dependent functions. In this case, from \eqqref{eq:varphi} we would have $i\exv{{O'}^*O'}=2\nu a_\phi$. Now assume that we have obtained a numerical solution to the linearized equations of motion \eqqref{eq:leom}, which is a formidable task in practice. Then we would be able to read off the coefficients $\delta\phi_v$ and $\delta\phi_s$. Since $\delta\phi_v$ in general includes contributions from sources other than $\delta\phi_s$, we cannot find  $\exv{{O'}^*O'}$ from simply taking the ratio of these coefficients. However, exploiting the linearity of \eqqref{eq:leom} enables us to find a solution which on the boundary only sources the scalar field fluctuations. We then have that all source fluctuations $\delta\Phi_s^I$ vanish except for $\delta\phi_s$, so that $\delta\phi_v=a_\phi\delta\phi_s$ in \eqqref{eq:deltaphiexp}. As a consequence, for this particular solution the correlator $\exv{{O'}^*O'}$ can be found from the ratio of $\delta\phi_v$ and $\delta\phi_s$ \cite{fn6}.

It is important to realize that the two-point function $\exv{{O'}^*O'}$ found by the procedure above does \textit{not} coincide with the intrinsic dynamics of the order parameter fluctuations. Instead, this two-point function will contain the full dynamics, and thus have poles for all the quasinormal modes of the system. We will now proceed with explaining what we mean by the intrinsic dynamics.
\subsection{Intrinsic dynamics} \label{ss:ID}
\begin{figure}[!t]
\subfloat[Intrinsic propagator\label{fig:intr}]
	{\includegraphics[trim = 3cm 6.5cm 3cm 1.6cm, clip=true, scale=.125]{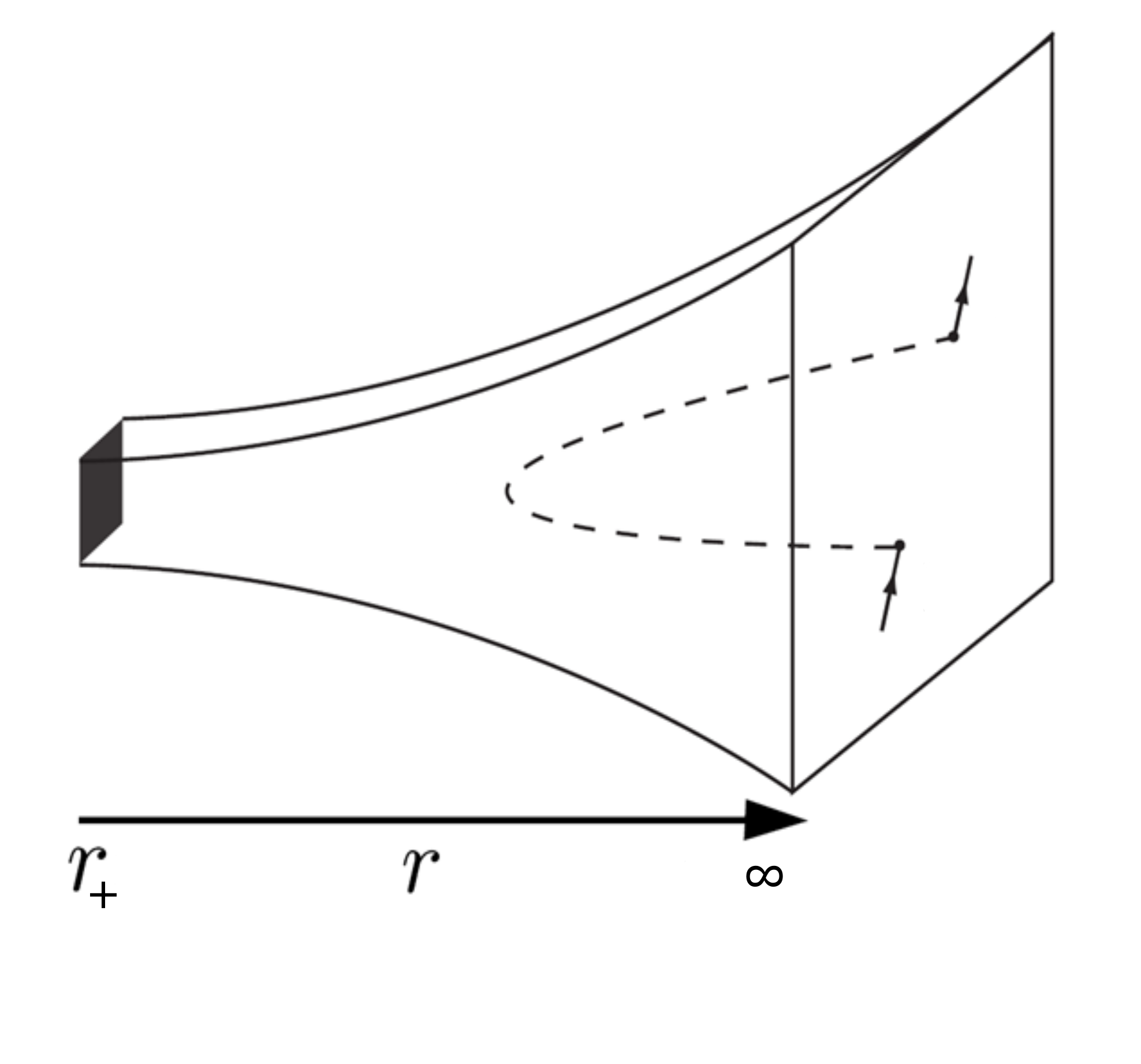}}

\subfloat[Full propagator\label{fig:full}]
	{\includegraphics[trim = 3cm 6.5cm 3cm 1.6cm, clip=true, scale=.125]{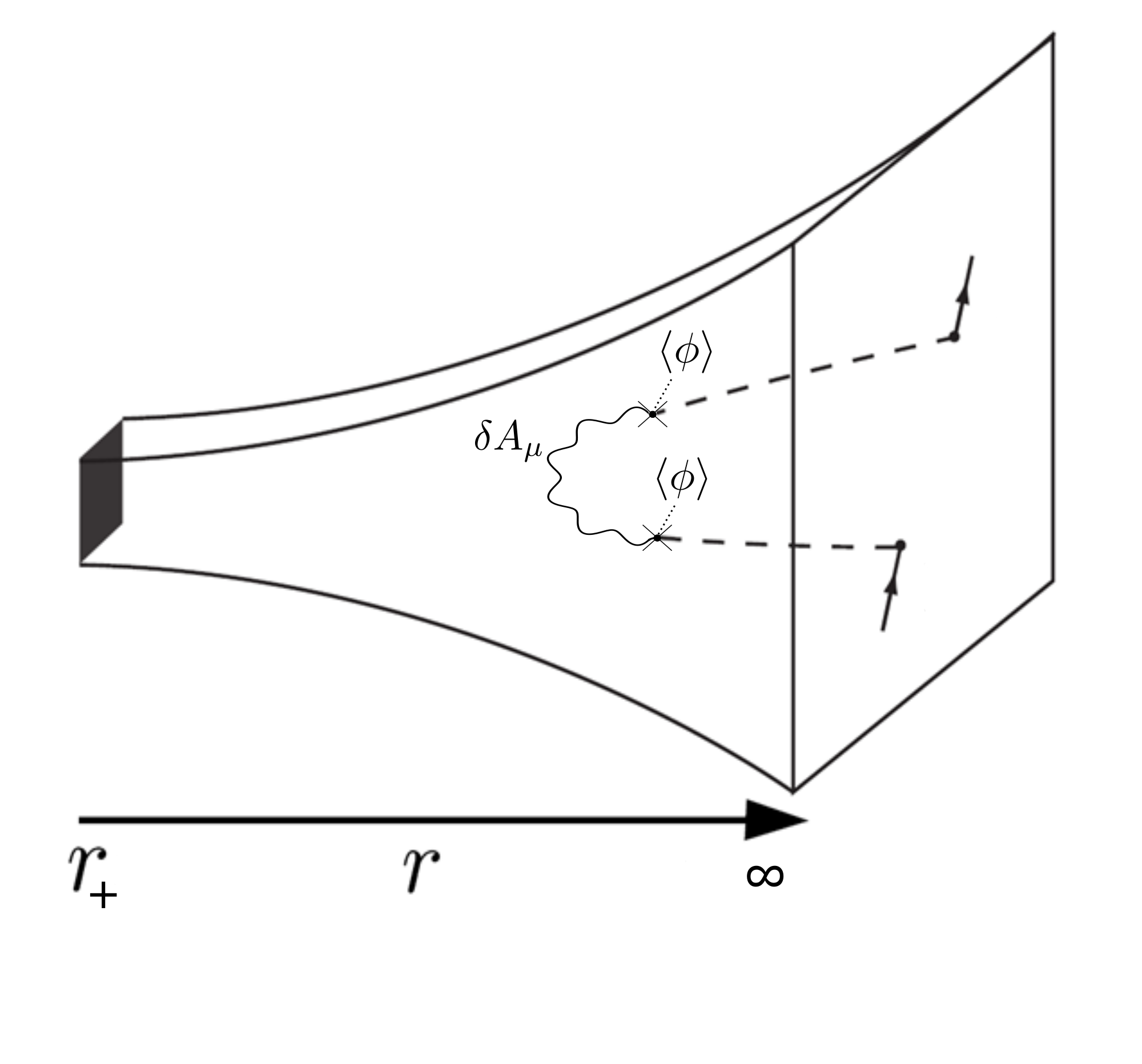}}
\caption{(a) Illustration of the intrinsic bulk-to-boundary propagator. The dashed curve denotes the scalar field fluctuations propagating into the bulk geometry. (b) Illustration of a contribution to the full bulk-to-boundary propagator. This shows the scalar field fluctuations interacting with gauge field fluctuations. These figures are adapted versions of Fig. 2 in \cite{Jacobs}.}
\label{fig:b2b} 
\end{figure}

The intrinsic dynamics of the order parameter fluctuations correspond to the dynamics that is \textit{uncoupled} from the other hydrodynamic fluctuations in the theory. This means that we ignore the coupling of the order parameter fluctuations to the other hydrodynamic degrees of freedom. In the bulk theory, the uncoupled dynamics can be found by simply setting $\delta A_\mu$ and $\delta g_{\mu\nu}$ to zero. The intrinsic retarded Green's function for the order parameter $G_{R,O}^{intr}$ is still given by the expression
\be{ \label{eq:retardedGprobe}
G^{intr}_{R,O}(\omega,\vec{k})=2\nu\frac{\partial\delta\phi_v}{\partial\delta\phi_s}.}
However, in contrast to the full retarded Green's function, the variation of $\delta\phi_{v}$ with respect to $\delta\phi_{s}$ is now found from the equation of motion for the uncoupled order parameter fluctuations, i.e.,
\be{ \label{eq:probeEOM}
\lr{D_\mu D^\mu - m^2}\delta\phi=0.}
This equation is found from the linearized equations of motion in \eqqref{eq:leom} by just putting the other fluctuations to zero. Basically, this corresponds to the bulk-to-boundary propagator shown in Fig. \ref{fig:intr} . Here, the scalar field fluctuations propagate into the bulk without coupling to the other fluctuations that are present there. As a result, we obtain the uncoupled dynamics of the order parameter fluctuations. In general, this intrinsic dynamics does not coincide with the full dynamics. Indeed, notice that a solution to \eqqref{eq:probeEOM} is in general \textit{not} a solution to the linearized equations of motion in \eqqref{eq:leom}. This is only the case when there is no coupling between the different fluctuations. This represents the way in which holography accounts for operator mixing in the boundary field theory. In a general theory where couplings are present, the quasinormal modes are thus modified by the coupling between the intrinsic modes. These quasinormal modes are found from the full propagator, which was discussed in the previous subsection. In Fig.  \ref{fig:full} we illustrate one of the contributions to the full propagator. Here, the scalar field fluctuations propagate into the bulk where they couple to the gauge field fluctuations that are present there. Similar contributions arise from the interactions with the fluctuations of the metric. The two-point vertices that are depicted as crosses in this figure are proportional to the background value $\exv{\phi}$, such that this contribution vanishes in the normal phase as expected. The quasinormal modes obtained from the full propagator would be the modes that can be directly observed if the system were experimentally realized. Nevertheless, the computation of the intrinsic order parameter dynamics can yield interesting information about the nature of the order parameter. The main reason for this is that a Ginzburg-Landau model for the order parameter would also describe the intrinsic dynamics. This should become clear in the following explanation of the physical meaning of intrinsic dynamics in the boundary theory. 

In the boundary theory, we could in principle calculate the full set of all coupled correlation functions, including for instance $\exv{{J'}^\mu {J'}^\nu}$, $\exv{{J'}^\mu O'}$, $\exv{{O'}^* O'}$, and $\exv{{T'}^{\mu\nu} O'}$, using the approach in the previous subsection. Here $J^\mu$ is the $U(1)$ current dual to the gauge field and $T^{\mu\nu}$ is the energy-momentum tensor dual to the metric $g_{\mu\nu}$. In terms of an effective action for the longitudinal dynamics, this is equivalent to a mode-coupling theory where the order parameter fluctuations $O'$ are coupled to the mass-density fluctuations $m'$ and the charge fluctuations $q'$. This is a consequence of the Ward identities for the various Green's functions, which follow from the conservation of charge, energy, and momentum.

To illustrate more clearly what this means, consider the effective action $S_{\text{eff}}$ of the field theory, expanded up to second order in the fluctuations $O'$, ${O'}^*$, $q'$, and $m'$. Naturally, the terms in the action that are linear in the fluctuations then give the equations of motion. For the mass density, this results in the conservation equation for the energy-momentum tensor $\partial_\mu \exv{T^{\mu\nu}}=0$. This equation gives rise to two longitudinal sound modes and two transverse diffusive modes. Furthermore, the charge density fluctuations lead to the conservation of the associated $U(1)$ current, i.e., $\partial_\mu \exv{J^\mu}=0$. This results in a diffusive mode for the intrinsic, i.e., uncoupled dynamics of the charge fluctuations. For the order parameter, the associated equation of motion can be written as $\delta\mathcal{L}_{GL}/\delta \exv{O}=0$ and its conjugate, where $\mathcal{L}_{GL}$ is a Ginzburg-Landau Lagrangian density, which we will discuss in the next sections.

Proceeding with the expansion, we can write the terms in the action that are of second order in the longitudinal fluctuations as
\be{
S_{\text{eff}}^{(2)}=-\frac{1}{2}\int\dd t\int\dd^3\vec{x}\,\vec{v}^\dagger G_R^{-1}\vec{v}.}
Here, we defined the Nambu-space fluctuation vector
\be{
\vec{v}\equiv \begin{pmatrix}
O'\\
\,{O'}^{*}\\
q'\\
m'
\end{pmatrix}.}
Furthermore, we can write the inverse retarded Green's function matrix $G_R^{-1}$ as
\be{ \label{eq:GRmatrix}
G_R^{-1}= \begin{pmatrix}
G_{R,O}^{-1} & \begin{array}{cc}\times & \times \\ \times & \times \end{array}\\
\begin{array}{cc}\times & \times \\ \times & \times \end{array} & ~\begin{array}{rl} G^{-1}_{R,q} & \times \\ \times & G^{-1}_{R,m} \end{array}
\end{pmatrix}.}
The inverse retarded Green's functions $G_{R,O}^{-1}$, $G_{R,q}^{-1}$, and $G_{R,m}^{-1}$ contain the intrinsic longitudinal modes associated with the continuity equations described above. Notice that $G_{R,O}^{-1}$ is a $2\times 2$ matrix and that the intrinsic Green's function in \eqqref{eq:retardedGprobe} corresponds to the upper-left component of the inverse of this matrix. In terms of a Dyson equation, we are basically computing the Green's function without the self-energy correction due to the interactions with the other modes. 

In general, the fluctuations of the order parameter will thus couple to the charge-density and mass-density fluctuations. These couplings are depicted as crosses in the matrix in \eqqref{eq:GRmatrix}. As a result, the modes corresponding to the retarded Green's function with the full dynamics, i.e., the inverse of the full matrix $G_R^{-1}$, are different from the modes associated with the intrinsic dynamics. Most importantly, the coupling between the diffusive mode associated with the charge-density fluctuations and the intrinsic mode associated with the order parameter result in two sound-like modes. This is for instance shown analytically in the probe limit in Ref. \cite{HerzogR}. We discuss this example in more detail at the end of this paper. However, in the remainder of this paper we will focus on the dynamics following from the Ginzburg-Landau Lagrangian density alone, so that we only need to study the intrinsic dynamics of the order parameter. Nonetheless, we have checked the consistency of our results with those in Ref. \cite{HerzogR}. In particular, using the full set of Green's functions obtained there for the coupled order-parameter and current fluctuations, we reconstructed the $3 \times 3$ inverse Green's function that appears in the effective action for the order parameter and charge-density fluctuations. The $2\times 2$ part corresponding to the order-parameter fluctuations in this inverse Green's function compares favorable to our results presented in Subsection \ref{ss:SP} below. We note that the analysis in Ref. \cite{HerzogR} is performed in the probe limit. Despite this, the comparison with our results is justified by the fact that the analysis is performed near the critical temperature, where backreaction effects are negligible.

In the next subsections, we will present our result for the intrinsic order parameter dynamics. These are obtained as follows. In Fourier space, the equation in \eqqref{eq:probeEOM} for the order parameter fluctuations can be written as 
\ba{ \label{eq:probeEOM2}
\delta\phi''&+\lr{\frac{f'}{f}+\frac{d-1}{r}-\frac{\chi'}{2}}\delta\phi'\nonumber\\&-\frac{m^2-(qA_t+\omega)^2\frac{e^\chi}{f}+\frac{|\vec{k}|^2}{r^2}}{f}\delta\phi=0,}
The conjugate equation holds for $\delta\phi^*$. Notice that this implies that $\delta\phi^*$ is not coupled to $\delta\phi$. This implies that $\partial\delta\phi_v/\partial\delta\phi^*_s=0$. As a consequence, we can always calculate \eqqref{eq:retardedGprobe} by first numerically solving \eqqref{eq:probeEOM2} and then computing the ratio of the resulting coefficients $\delta\phi_{v}$ and $\delta\phi_{s}$. When considering the full dynamics this is not the case, e.g. due to the coupling of both $\delta\phi$ and $\delta\phi^*$ to $\delta A_\mu$. As usual, we require infalling boundary conditions at the horizon, corresponding to the \textit{retarded} Green's function. Naturally, besides depending on frequency and momentum, this two-point function depends on the background parameters, i.e., on $q$, $m^2$ and $T/\mu$.
\subsection{The normal phase} \label{ss:NP}
To obtain the intrinsic retarded Green's function in the normal phase, we must solve \eqqref{eq:probeEOM2} with $\chi=0$, and with $A_t$ and $f$ given by Eqs. \eqref{eq:Atnormal} and \eqref{eq:fnormal} respectively. We have done so numerically. Using the numerical solution, we obtain the retarded Green's function by using Eqs. \eqref{eq:deltaphiexp} and \eqref{eq:retardedGprobe}. Notice that this coincides with the full retarded Green's function in this case, because in the normal phase the fluctuations are decoupled.

Approaching the transition temperature from above, the physics can be described with a time-dependent Ginzburg-Landau model. This can be represented by the action
\ba{ 
S=&-\int\dd t\int\dd^3\vec{x}\nonumber \\ &\times\lr{iaO^*\partial_t O+\gamma|\vec{\nabla}O|^2+\alpha|O|^2+\frac{\beta}{2}|O|^4},\label{eq:LGaction2}}
which incorporates the result of \eqqref{eq:expv}. The first two terms of the integrand capture the long-wavelength and low-frequency behavior of the order parameter. Like $\alpha$ and $\beta$, the coefficients $\gamma$ and $a$ depend on the temperature. The coefficient $a$ is complex, since the system shows dissipation of the order parameter. This implies that the imaginary part of $a$ is negative. Physically, dissipation occurs as a consequence of temperature fluctuations, which can cause the fermion pairs to break up. From the above action, we obtain the retarded Green's function of $O$ in this model from the part of the action that is quadratic in the order parameter fluctuations, which we can subsequently compare with the retarded Green's function obtained holographically. Thus, this is a tree-level calculation, even though AdS/CFT should provide us with the \textit{full} partition function of the dual field theory. The reason is that contributions of higher orders in the fluctuations are suppressed by the large-$N$ limit, implicit in the AdS/CFT correspondence, and loop diagrams come with factors of $1/N$. This claim is motivated by the mean-field result for the critical exponent in Fig. \ref{fig:OTmu} , which does not change when taking into account only Gaussian fluctuations.

\begin{figure}[!t]
\subfloat[$\alpha_0$\label{fig:Coefalpha}]
	{\includegraphics[trim = 2.4cm 0.1cm 0cm 0cm, clip=true, scale=.31]{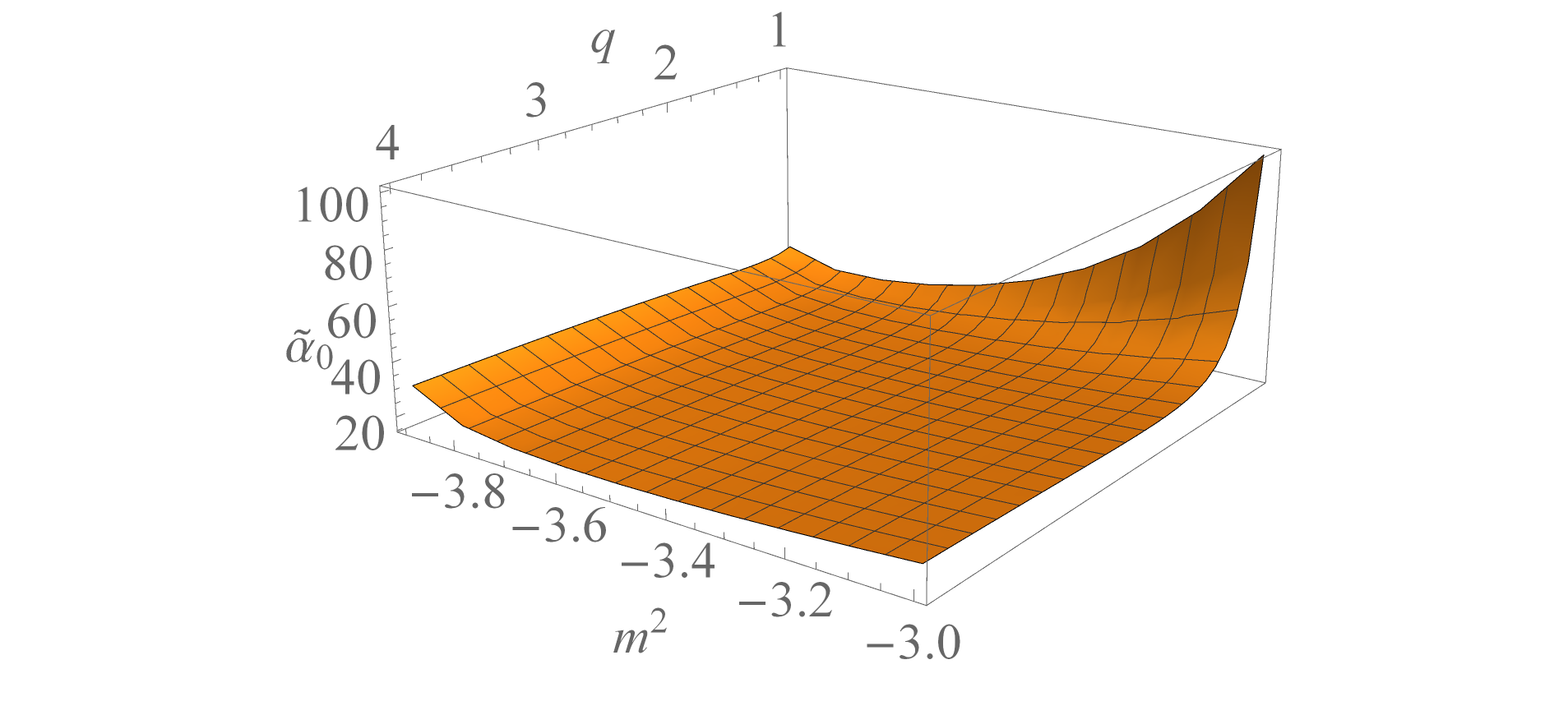}}

\subfloat[$\gamma$\label{fig:Coefgamma}]
	{\includegraphics[trim = 5cm 0cm 0cm 0cm, clip=true, scale=.34]{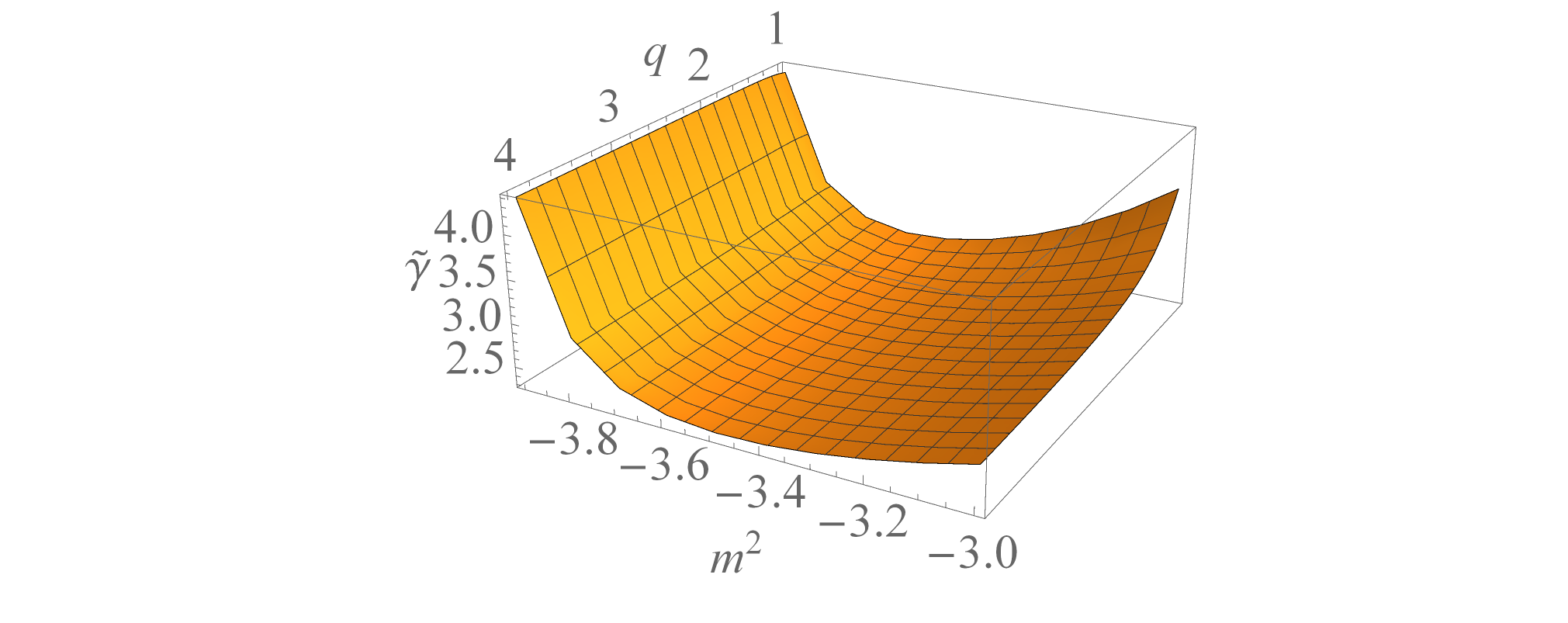}}

\subfloat[Re$(a)$\label{fig:CoefRea}]
	{\includegraphics[trim = 0cm 0cm 0cm 0cm, clip=true, scale=.29]{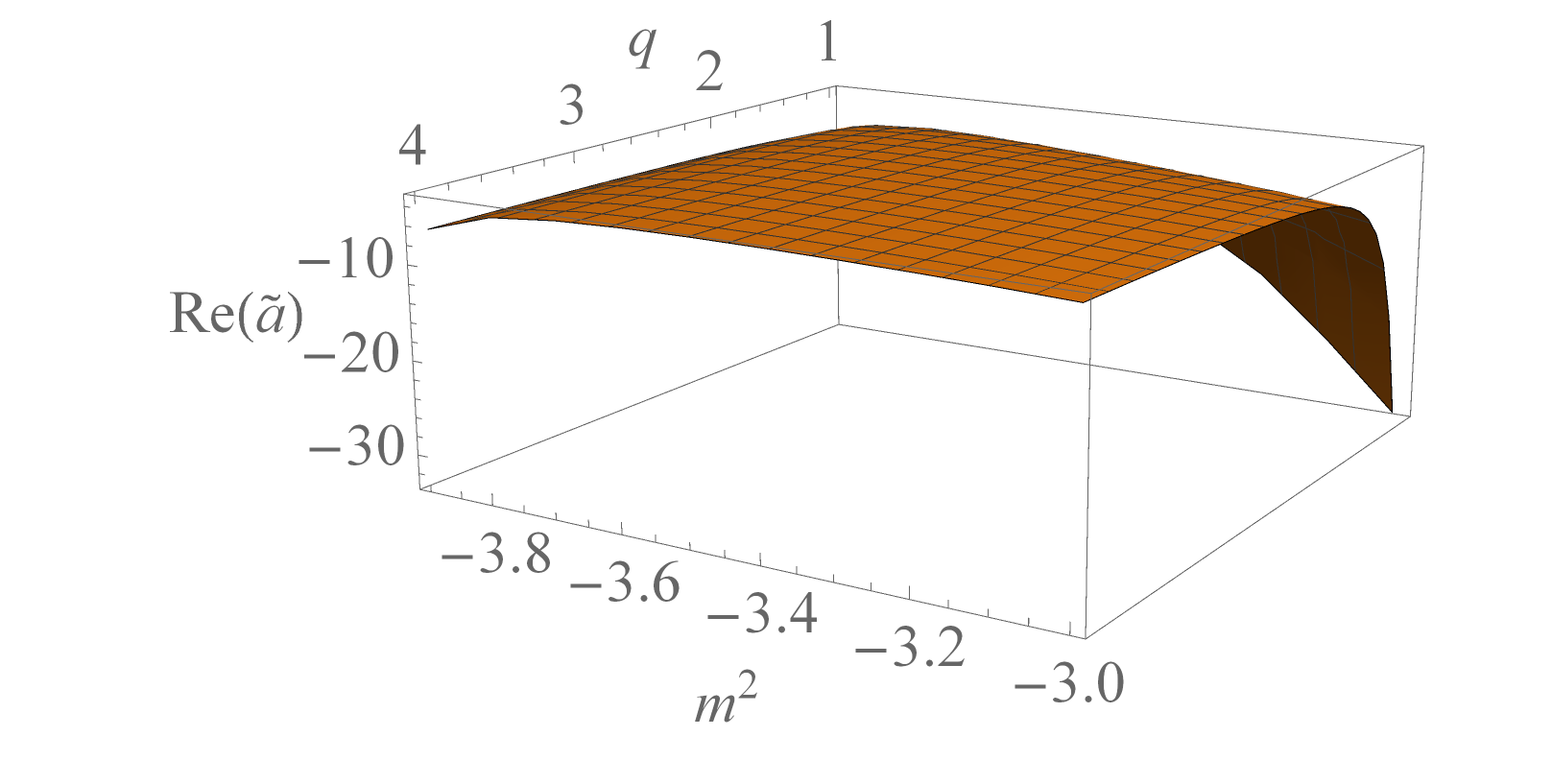}}

\subfloat[Im$(a)$\label{fig:CoefIma}]
	{\includegraphics[trim = 0cm 0.1cm 0cm 0cm, clip=true, scale=.28]{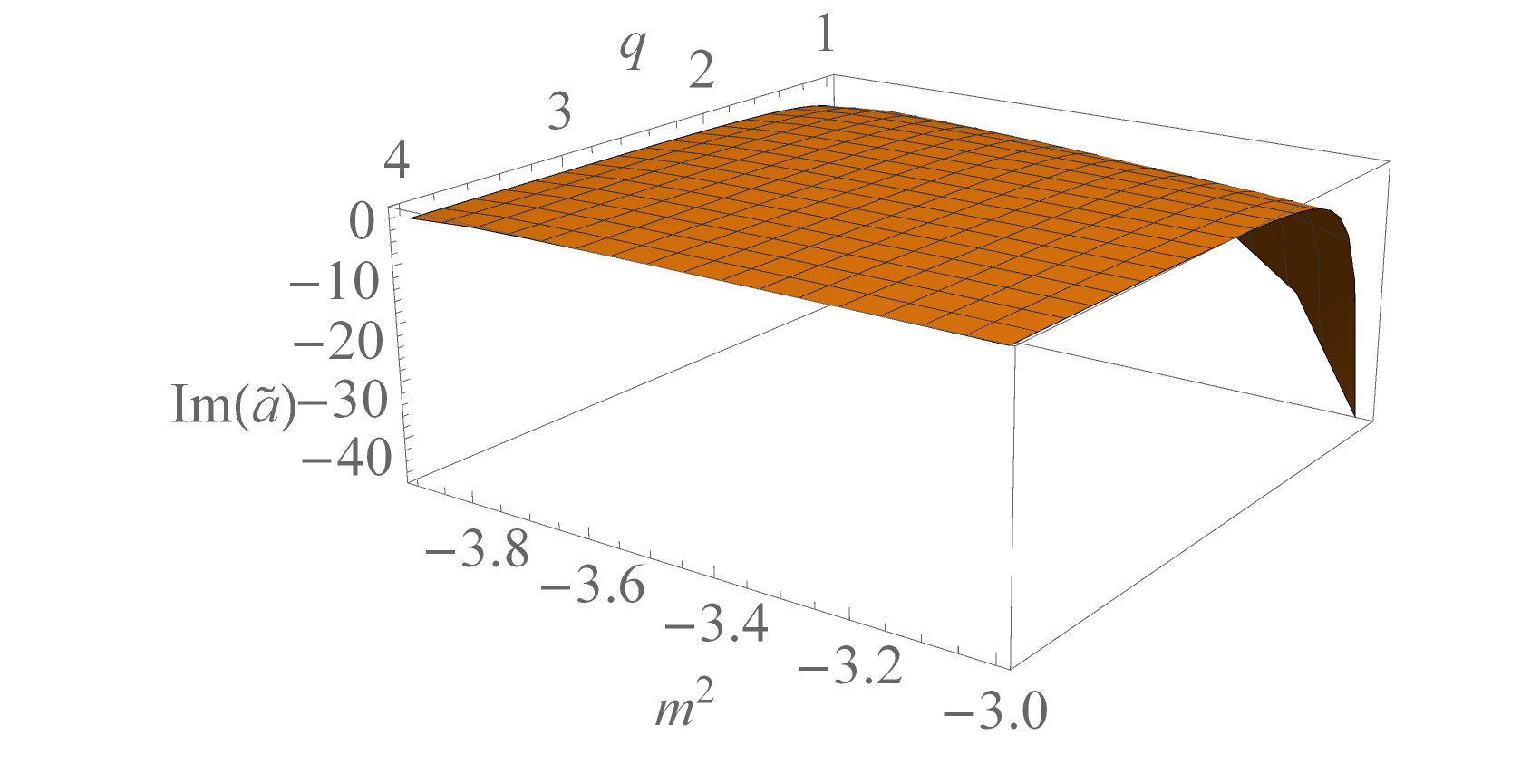}}
\caption{The parameters in the Ginzburg-Landau action as a function of $q$ and $m^2$. The parameter $\beta$ follows from Fig. \ref{fig:propTc} . The tildes above the parameters imply that they are scaled with appropriate powers of $\mu$ to make them dimensionless.}
\label{fig:coefdata}
\end{figure}

Since the order parameter has a vanishing expectation value in the normal phase, the part of the Ginzburg-Landau action that is quadratic in the order parameter fluctuations $O'\equiv O-\exv{O}=O$ is given by
\ba{ \label{eq:LGactionabove}
S_{\text{quad}}=&\frac{-1}{(2\pi)^4}\int\dd\omega\int\dd^{3}\vec{k}\nonumber \\ &\times {O'}^{*}(\omega,\vec{k})[a\omega+\gamma|\vec{k}|^2+\alpha_0(T-T_c)]O'(\omega,\vec{k})}
near the transition temperature $T_c$. From this we obtain that the two-point function is given by
\be{ \label{eq:tpnp}
i\left<{O'}^{*}O'\right>(\omega,\vec{k})=\frac{1}{a\omega+\gamma|\vec{k}|^2+\alpha_0(T-T_c)}.}
This result should hold for small frequencies and momenta, i.e., $\omega\ll \mu$ and $|\vec{k}|\ll \mu$, since the Ginzburg-Landau action in \eqqref{eq:LGaction2} only contains the leading orders of the gradient expansion. Comparing the above expression to our numerical results, we can determine the coefficients $\alpha_0$, $a$, and $\gamma$ near $T_c$. The result is shown in Fig. \ref{fig:coefdata} . Together with the result from Fig. \ref{fig:propTc} in the previous section, we then obtain all the coefficients in \eqqref{eq:LGaction2} near the critical temperature.

Although quantitatively, the results clearly depend on $m^2$ and $q$, the qualitative physics does not seem very different for different values of these parameters. Therefore we will restrict the following discussion of the retarded Green's function to the fixed values $q=3$ and $m^2=-3.5$.
\begin{figure}[!t]
\includegraphics[width=.95\linewidth,clip]{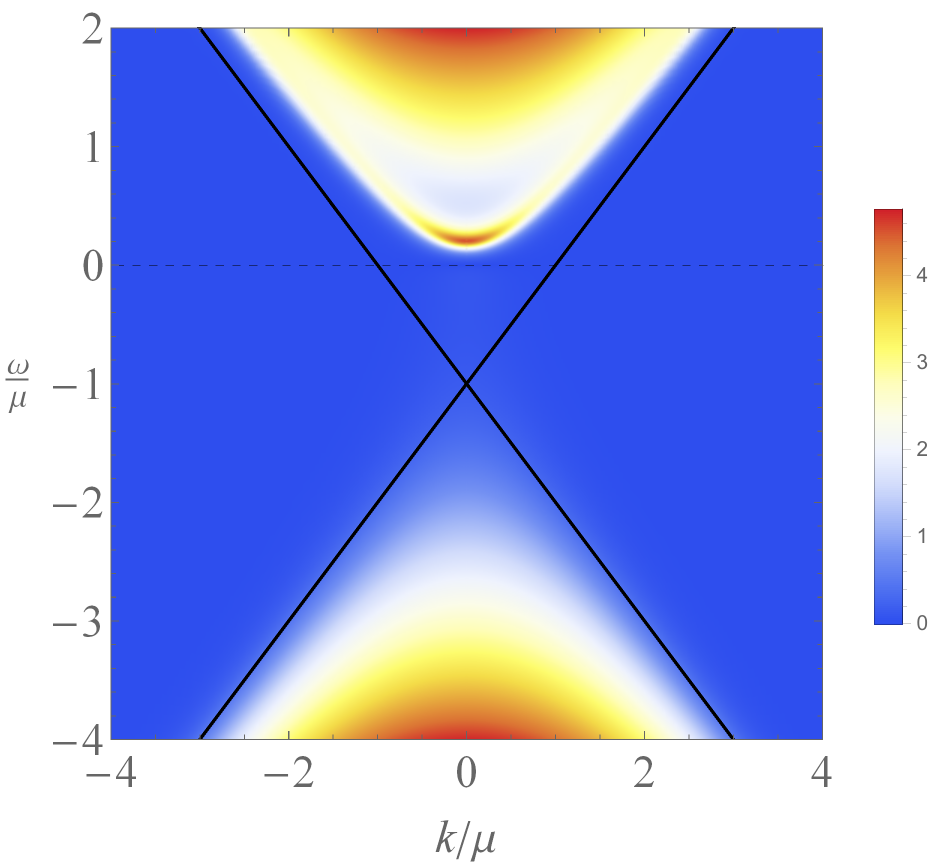}
\caption{\label{fig:sp152}The spectral function for $T = 1.5 T_c$, $q=3$, and $m^2=-3.5$. Here and in all following plots of the spectral functions, we have shown the absolute value of the spectral functions and divided by a factor $\mu^{2\nu}$ to make them dimensionless.}
\end{figure}

From the retarded Green's function $ i\left<{O'}^{*}O'\right>$, we can obtain the spectral function
\be{
\rho(\omega,\vec{k})=\frac{1}{\pi}\,\text{Im}\big[ i\left<{O'}^{*}O'\right>(\omega,\vec{k})\big].}
This is an interesting quantity, as it yields the dispersion relations of the modes accessible to the intrinsic order parameter dynamics as well as the corresponding lifetimes. In Fig. \ref{fig:sp152} this quantity is shown at the temperature $T=1.5T_c$. Here we have plotted the absolute value of the spectral function, noting that the spectral function itself is negative for $\omega<0$. Moreover, we have exploited rotational invariance to fix the direction of $\vec{k}$, such that $k$ denotes the component in that direction. Naturally the spectral function is symmetric in $k$. In accordance with the Green's function in \eqqref{eq:tpnp} obtained in the Ginzburg-Landau model, we see that the spectral function vanishes for $\omega = 0$, since $\alpha$, $\beta$, and $\gamma$ are real coefficients. Furthermore, for small $\omega$ and $\vec{k}$, we also see a quadratic dispersion as predicted by \eqqref{eq:tpnp}, which is shifted upward from $\omega=0$ since $\alpha$ is nonzero. When $\omega$ is large compared to $\mu$ and $T_c$, we recover the spectral function from pure AdS, which is given by (see e.g. Ref. \cite{SonStarinets})
\be{ \label{eq:spfAdS}
\rho_{\text{AdS}}(\omega,\vec{k})=\frac{2\nu}{\pi}\text{Im}\lr{\frac{\sqrt{-\omega^2+|\vec{k}|^2}}{2}}^{2\nu}\frac{\Gamma(-\nu)}{\Gamma(\nu)},}
where $\nu= \sqrt{d^2+4m^2}/2$ as before and $\Gamma$ denotes the gamma function. In Fig. \ref{fig:sp152} we observe that the spectral weight fills the light cone which is shifted down by the chemical potential, i.e., $|\omega+\mu|=|k|$. This cone is shown in black in the figure, where the momentum space domain is taken small enough such that shift is still visible.

\begin{figure}[!t]
\includegraphics[width=.95\linewidth,clip]{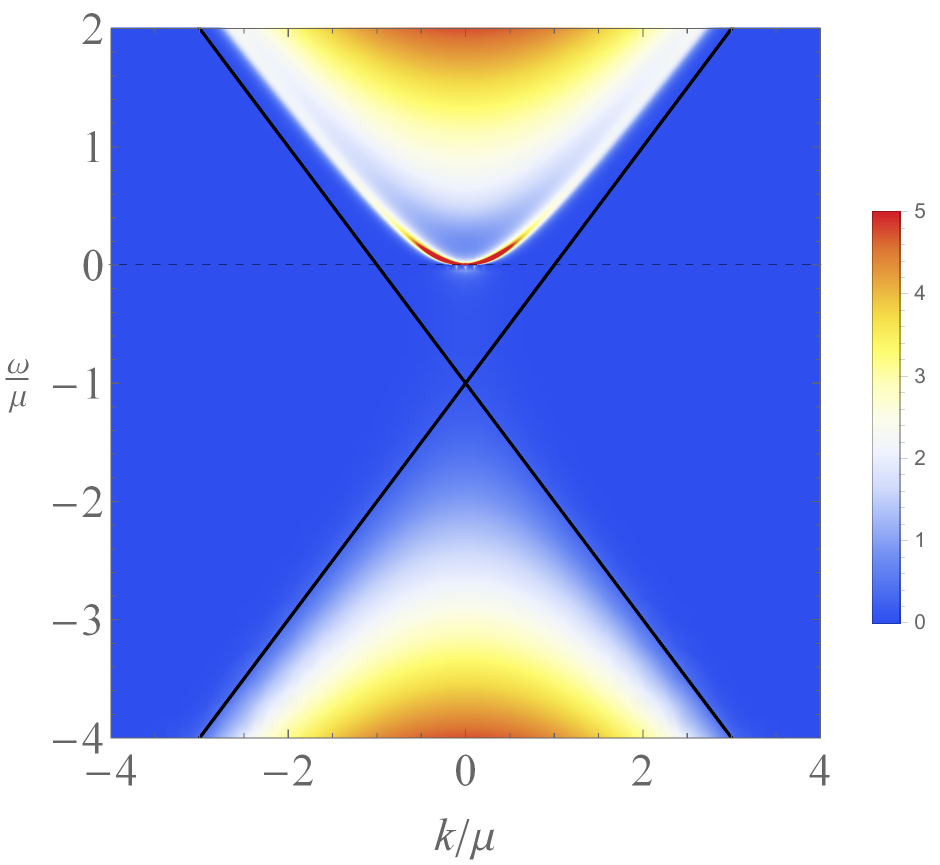}
\caption{\label{fig:sp2}The spectral function as $T\rightarrow T_c$ for $q=3$ and $m^2=-3.5$.}
\end{figure}

In Fig. \ref{fig:sp2} , the spectral function for $T=T_c$ is shown. As before, we can distinguish two regimes here, namely the UV regime in which we recover the AdS result and the IR regime in which the physics can be described with the Ginzburg-Landau model. In the latter regime we again see the quadratic dispersion predicted by \eqqref{eq:tpnp}, which gives a peak centered at $\omega_{\text{peak}}=-\gamma |\vec{k}|^2\text{Re}(a)/|a|^2$ since now $\alpha= 0$ in \eqqref{eq:LGaction}.

Notice that in Ref. \cite{SchalmZaanen}, a similar approach to the retarded Green's function is given for the zero-momentum case. Although the UV results are different because Ref. \cite{SchalmZaanen} uses alternative quantization, the IR results are comparable, i.e., in both cases the results can be described in the low-frequency limit with the Ginzburg-Landau model.
\subsection{The superconducting phase} \label{ss:SP}
Using the same method as in the normal phase, we can determine the intrinsic Green's function in the superconducting phase numerically. For this purpose, we must solve \eqqref{eq:probeEOM2} again, where this time $f$, $\chi$, and $A_t$ are the numerical functions discussed in the previous chapter.

To compare the result with the Ginzburg-Landau model as before, we first expand $O$ around its expectation value as $O=\exv{O}+O'$. Upon doing so, the quartic term in \eqqref{eq:LGaction} yields
\ba{
\beta|O|^4&=\beta\exv{O}^2\lr{4|O'|^2+{O'}^{*}{O'}^{*}+O'O'}+\dots\nonumber \\&=-\alpha\lr{4|O'|^2+{O'}^{*}{O'}^{*}+O'O'}+\dots,}
where the dots denote terms which are not quadratic in the fluctuations and where we used that $\exv{O}$ is real and given by \eqqref{eq:expv}. Using this, we can write the part of the Ginzburg-Landau action in momentum space that is quadratic in the fluctuations as
\ba{ \label{eq:quadacsp}
S_{\text{quad}}=&\frac{-1}{(2\pi)^4}\int\dd \omega\int\dd^{3} \vec{k}\nonumber \\ &\times\bigg\{ {O'}^{*}(\omega,\vec{k})\lr{a \omega+\gamma|\vec{k}|^2-\alpha}O'(\omega,\vec{k})\nonumber \\& \qquad -\frac{\alpha}{2}\big[{O'}^{*}(-\omega,-\vec{k}){O'}^{*}(\omega,\vec{k})\nonumber \\ &\qquad\qquad+O^{'}(\omega,\vec{k})O^{'}(-\omega,-\vec{k})\big]\bigg\}.}
The intrinsic retarded Green's function, defined by
\be{
G_R(\omega,\vec{k})\equiv i\begin{pmatrix}
\exv{{O'}^{*} O'} && \exv{{O'}^{*}{O'}^{*}} \\
\exv{O'O'} && \exv{O'{O'}^{*}} \\
\end{pmatrix},}
can then be found from
\ba{ \label{eq:GRreadoff}
S_{\text{quad}}= \frac{1}{2}\frac{-1}{(2\pi)^4}\int\dd \omega\int\dd^{3} \vec{k}\ &\begin{bmatrix}
{O'}^{*}(\omega,\vec{k}) &&
O'(-\omega,-\vec{k}) \\
\end{bmatrix}\nonumber \\&\times G_R^{-1} \begin{bmatrix}
O'(\omega,\vec{k}) \\
{O'}^{*}(-\omega,-\vec{k}) \\
\end{bmatrix}.}
\begin{widetext}
From this we obtain the intrinsic two-point function $\exv{{O'}^{*} O'}$. Using \eqqref{eq:quadacsp}, this then yields
\be{ \label{eq:tpsp}
i\exv{{O'}^{*}O'}=\frac{a^*\omega-\gamma |\vec{k}|^2+\alpha}{|a|^2\omega^2-\gamma|\vec{k}|^2\lr{\gamma|\vec{k}|^2-2\alpha}-2i\, \text{Im}(a)\omega\lr{\gamma |\vec{k}|^2-\alpha}}.}
\end{widetext}
This expression for the two-point function allows us to make several predictions. First of all, for $\omega=0$ and close enough to the critical point such that we can still approximate $\alpha\approx \alpha_0(T-T_c)$, we can write
\be{ \label{eq:apprx}
\gamma|\vec{k}|^2i\exv{{O'}^{*}O'}\approx\frac{\gamma |\vec{k}|^2-\alpha_0(T-T_c)}{\gamma|\vec{k}|^2-2\alpha_0(T-T_c)}.}
Given some small but nonzero $\alpha$, the above quantity approaches $1$ as a function of $k$ in the regime where $\gamma|\vec{k}|^2\gg |\alpha|$, but $|\vec{k}|\ll \mu$ so that the long-wavelength limit is still valid. This just shows that $\exv{{O'}^{*}O'}$ is continuous at $T=T_c$. In contrast, in the regime where $\gamma|\vec{k}|^2\ll |\alpha|$, the above quantity approaches $1/2$. Hence we should easily be able to distinguish both regimes. However, in both regimes the numerics show that $i\exv{{O'}^{*}O'}\approx 1/(\gamma|\vec{k}|^2)$. We can therefore conclude that upon lowering the temperature below $T_c$, the Green's function in \eqqref{eq:tpsp} does not reproduce the holographic results.
\begin{figure}[!t]
\includegraphics[width=.95\linewidth,clip]{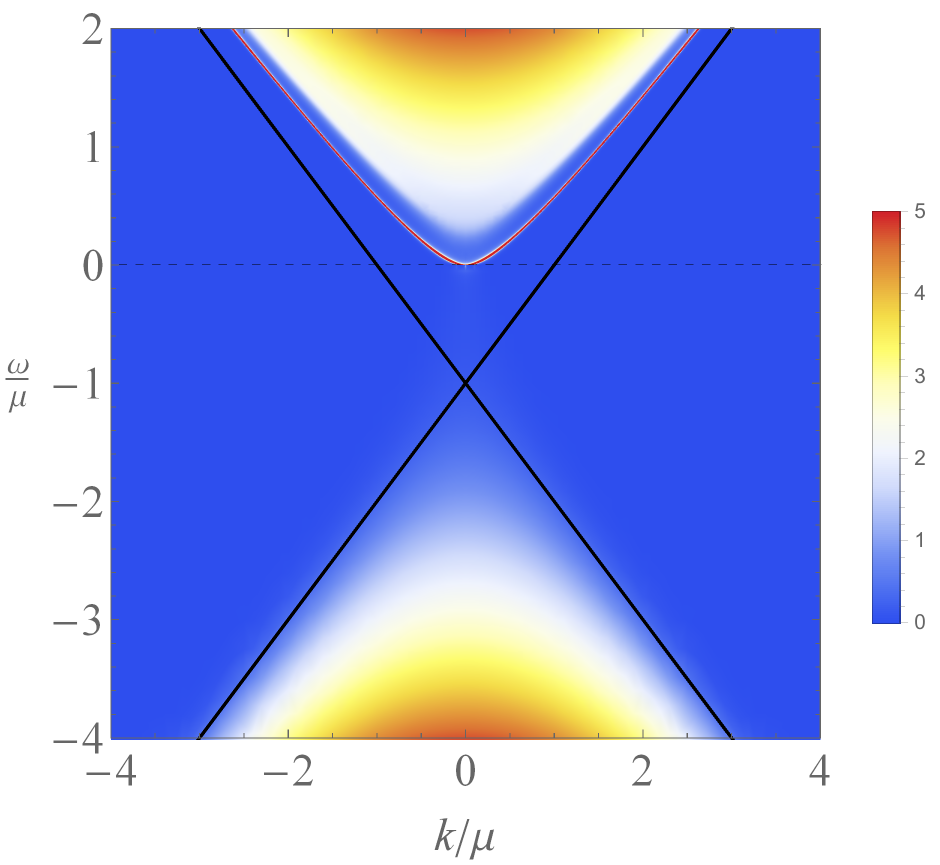}
\caption{\label{fig:sp4}The spectral function for $T= 0.5T_c$, $q=3$, and $m^2=-3.5$.}
\end{figure}

Moreover, the Ginzburg-Landau model predicts strongly overdamped sound-like modes. This can be seen from the poles in \eqqref{eq:tpsp}, which are located at
\ba{
\omega=\frac{1}{|a|^2}\bigg[&i \text{Im}(a)(\gamma |\vec{k}|^2-\alpha)\nonumber \\ &\pm \sqrt{\text{Re}(a)^2\gamma |\vec{k}|^2(\gamma |\vec{k}|^2-2\alpha)-\alpha^2\text{Im}(a)^2}\bigg].}
In the limit $\gamma |\vec{k}|^2\ll |\alpha|$, the poles are thus purely imaginary as long as
\be{ \frac{\text{Im}(a)^2}{\text{Re}(a)^2}\gtrsim \frac{2\gamma |\vec{k}|^2}{|\alpha|}.}
As expected for the retarded Green's function, the imaginary part of the poles is always negative. Nevertheless, even within this limit the spectral functions corresponding to the Ginzburg-Landau model contain peaks at both positive and negative frequencies. These peaks sharpen up at lower temperatures and at zero temperature, where $\text{Im}(a)=0$, become the two long-lived Anderson-Bogoliubov sound modes with the dispersion $\omega=\pm\sqrt{2\gamma|\alpha|}|\vec{k}|/|a|$ in the long-wavelength limit. Neither the two peaks in the strongly overdamped regime nor the sound modes are observed in our holographic results. The peak we do observe in the numerical spectral functions is at long-wavelengths positioned only at the positive frequency $\omega_{peak}=-\gamma |\vec{k}|^2\text{Re}(a)/|a|^2$. This is seen, for example, in the intrinsic spectral function in Fig. \ref{fig:sp4} where $T=T_c/2$. Note that since we have broken Lorentz symmetry by adding a nonzero chemical potential, we might naively expect that there will be Goldstone modes with a quadratic dispersion, also known as type-II Goldstone bosons. In our time-dependent Ginzburg-Landau theory, and in accordance with the counting rules discussed in Refs. \cite{Chadha,Amado,Mezz}, this however does not occur, due to the coupling between the phase and the amplitude of the order parameter which gives rise to a Goldstone mode with linear dispersion when $\text{Im}(a)=0$. Furthermore, beyond the leading order of the long-wavelength limit superconductors typically have a Higgs mode, which is not observed in the spectral functions. It therefore seems that our results from holography are inconsistent with the abovementioned Ginzburg-Landau model. We therefore proceed by presenting an alternative model that does describe the numerical results we have obtained.
\subsection{The large-$N$ Ginzburg-Landau model} \label{ss:LN}
As mentioned in the previous section, the action of the bulk theory is proportional to a dimensionless constant related to the number of species in the theory. Therefore, we propose that the dual field theory can be described by an effective theory of $N$ complex order parameters, one of which will acquire a nonzero expectation value below the critical temperature. Our gravitational dual still contains only one complex field charged under $U(1)$. Hence, using holography we cannot describe each order parameter on its own, i.e., the gravitational theory does not contain dual fields to all these order parameters. Rather, the scalar field in the bulk is only dual to a specific combination of these order parameters, similar to a single-trace operator in a Yang-Mills field theory.

Thus, instead of the usual Ginzburg-Landau model with a complex scalar as an order parameter, we introduce a modified Ginzburg-Landau model where the order parameter is a complex $N$-component vector with components that we denote by $O_n$. The action in Fourier space then reads
\ba{ \label{eq:modaction}
S=&\frac{-1}{(2\pi)^4}\int\dd t\int\dd^{3} \vec{x}\nonumber \\& \times \sum_{n=1}^N \Bigg(iaO_n^*\partial_t O_n+\gamma|\nabla O_n|^2+\alpha|O_n|^2\nonumber \\ &\qquad\qquad+\frac{\beta}{2N}|O_n|^2\sum_{m=1}^N |O_m|^2\Bigg).}
Here $a$ is again a complex coefficient whereas the other coefficients are still real. All coefficients depend of course on the temperature $T$ and on $q$ and $m^2$. Choosing the vacuum expectation value to be real and along the first component then yields $\exv{O_{i\neq 1}}=0$ and
\be{ \label{eq:exvN}
\exv{O_1}=\exv{O_1^*}=\sqrt{-\frac{\alpha N}{\beta}}}
below $T_c$. Hence again, a symmetry gets spontaneously broken below the transition temperature. However, rather than a simple breaking of a $U(1)$ symmetry, this time a $U(N)$ symmetry gets broken to $U(N-1)$.

We proceed by studying the intrinsic dynamics of the order parameter fluctuations in this model. Above $T_c$, the expectation value of the order parameter vanishes and the part of the action quadratic in the fluctuations $O_n'$ reads
\begin{align}
S_{\text{quad}}=\frac{-1}{(2\pi)^4}\int\dd \omega\int\dd^{3} \vec{k}\nonumber \\  \times\sum_{n=1}^N {O'_{n}}^{*}(\omega,\vec{k})&\lr{a \omega+\gamma|\vec{k}|^2+\alpha}O'_{n}(\omega,\vec{k}).
\end{align}
We read off the retarded Green's function $G_R(\omega,\vec{k})$ using
\ba{
S_{\text{quad}}= &\frac{1}{2}\frac{-1}{(2\pi)^4}\int\dd \omega\int\dd^{3} \vec{k} \nonumber \\ &\times  \begin{bmatrix}
{O'_{1}}^{*}(\omega,\vec{k}) \\
O_1'(-\omega,-\vec{k}) \\
\vdots \\
{O'_{N}}^{*}(\omega,\vec{k}) \\
O_N'(-\omega,-\vec{k}) \\
\end{bmatrix}^T\textbf{G}_R^{-1}\begin{bmatrix}
O'_1(\omega,\vec{k}) \\
{O'_{1}}^{*}(-\omega,-\vec{k}) \\
\vdots \\
O'_N(\omega,\vec{k}) \\
{O'_{N}}^{*}(-\omega,-\vec{k}) \\
\end{bmatrix}.}
The result is then the $2N\times 2N$ matrix given by
\be{
\textbf{G}_R(\omega,\vec{k})=I_{N}\otimes \begin{pmatrix}
					\frac{1}{a\omega+\gamma|\vec{k}|^2+\alpha} & 0 \\ 0 & \frac{1}{-a^* \omega+\gamma|\vec{k}|^2+\alpha}
				\end{pmatrix},}
where $I_N$ denotes the $N\times N$ identity matrix. This implies that
\be{
i\exv{{O'_{i}}^{*}O'_i}=\frac{1}{a\omega+\gamma|\vec{k}|^2+\alpha}}
for all $i\in\{1,\dots,N\}$.

Below $T_c$, we expand the order parameter around its expectation value as $O_n=\exv{O_n}+O_n'$. The quartic term in the action \eqref{eq:modaction} then yields
\begin{align}
\frac{\beta}{2N}\sum_{n=1}^N\sum_{m=1}^N&O_n^*O_n O_m^*O_m \nonumber \\
=\frac{\beta}{2N}\sum_{n=1}^N\sum_{m=1}^N&\Big( 2|\exv{O_m}|^2{O'_{n}}^{*}O'_n+2\exv{O^*_n}\exv{O_m} {O'_{n}}^{*}O'_m \nonumber \\  +&\left.\exv{O_n}\exv{O_m} {O'_{n}}^{*}{O'_{m}}^{*}+\exv{O_n^*}\exv{O_m^*} O_n^{'}O_m^{'}\right) \nonumber  \\+&\dots\nonumber \\
=\sum_{n=1}^N(-\alpha {O'_{n}}^{*}&O'_n)-\alpha{O'_{1}}^{*}O'_1-\frac{\alpha}{2}\lr{{O'_{1}}^{*}{O'_{1}}^{*}+O'_1O'_1}\nonumber \\ +\dots
\end{align}
where the dots denote terms that are not quadratic in the fluctuations and where we used \eqqref{eq:exvN}. The part of the action quadratic in the fluctuations then becomes
\begin{align}
S_{\text{quad}}=&\frac{-1}{(2\pi)^4}\int\dd \omega\int\dd^3 \vec{k} \nonumber \\ &\times \Bigg\{\sum_{n=1}^N  {O'_{n}}^{*}(\omega,\vec{k})\lr{a \omega+\gamma|\vec{k}|^2-\alpha \delta_{n,1}}O_{n}'(\omega,\vec{k}) \nonumber \\ &\qquad-\frac{\alpha}{2}\big[{O'_{1}}^{*}(-\omega,-\vec{k}){O'_{1}}^{*}(\omega,\vec{k})\nonumber \\ &\qquad\qquad+O^{'}_{1}(\omega,\vec{k})O^{'}_{1}(-\omega,-\vec{k})\big]\Bigg\}.
\end{align}
This yields the intrinsic Green's function matrix
\be{
\textbf{G}_R= \begin{bmatrix}
\textbf{G}_{R,1}& \underline{\underline{0}} \\
\underline{\underline{0}} & I_{N-1}\otimes \begin{pmatrix}
					\frac{1}{a\omega+\gamma|\vec{k}|^2} & 0 \\ 0 & \frac{1}{-a^* \omega+\gamma|\vec{k}|^2}
		\end{pmatrix}
\end{bmatrix},\label{eq:NGR}}
where the part
\begin{widetext}
\be{
\textbf{G}_{R,1}= \frac{1}{|a|^2\omega^2-\gamma|\vec{k}|^2\lr{\gamma|\vec{k}|^2-2\alpha}-2i\, \text{Im}(a)\omega\lr{\gamma |\vec{k}|^2-\alpha}}\begin{pmatrix}
					a^*\omega-\gamma|\vec{k}|^2+\alpha & \alpha \\ \alpha & -a\omega - \gamma|\vec{k}|^2+\alpha
	\end{pmatrix}}
coincides with the intrinsic Green's function derived from the $U(1)$ Ginzburg-Landau model. We then obtain the two-point function
\be{
i\exv{{O'_{1}}^{*}O'_1}=\frac{a^*\omega-\gamma|\vec{k}|^2+\alpha}{|a|^2\omega^2-\gamma|\vec{k}|^2\lr{\gamma|\vec{k}|^2-2\alpha}-2i\, \text{Im}(a)\omega\lr{\gamma |\vec{k}|^2-\alpha}},}
\end{widetext}
whereas for $i\neq 1$ we obtain
\be{
i\exv{{O'_{i}}^{*}O'_i}=\frac{1}{a\omega+\gamma|\vec{k}|^2}.}
The $N-1$ additional two-point functions all describe transverse modes. From \eqqref{eq:NGR} we note that there are $2N-1$ massless modes in total: $N-1$ doubly degenerate quadratic Goldstone modes from the lower $(N-1)\times (N-1)$ block and one strongly overdamped mode in the 2$\times$2 block. This is consistent with Goldstone's theorem, as $2N-1$ symmetries are broken upon breaking the $U(N)$ symmetry to a $U(N-1)$ symmetry.

Now, we still need a relation between the order parameter $O$ in the holographic superconductor and the $O_i$ in our large-$N$ Ginzburg-Landau model. We propose that
\be{ \label{eq:Odef}
O=\frac{1}{\sqrt{N}}\sum_{i=1}^NO_i.}
We think of the $O_i$ as order parameters for $N$ fermion species. With this definition for $O$, it then follows from \eqqref{eq:exvN} that we indeed get the expectation value given by \eqqref{eq:expv}. Moreover, we obtain the two-point function
\ba{ 
\exv{{O'}^{*}O'}&=\frac{1}{N}\sum_{i=1}^N\exv{{O'_{i}}^{*}O'_{i}}\nonumber \\ &=\frac{N-1}{N}\frac{-i}{a\omega+\gamma|\vec{k}|^2}+\frac{1}{N}\exv{{O'_{1}}^{*}O'_1}\nonumber \\ &\approx\frac{-i}{a\omega+\gamma|\vec{k}|^2}\label{eq:largeNtpf}}
where we took the large-$N$ limit in the last step.

The behavior of the two-point function near $T_c$ is consistent with our findings. In particular, it explains why we cannot distinguish between the two momentum regimes discussed below \eqqref{eq:apprx}. Moreover, we can now see what happens if we explore regimes of temperatures even further below $T_c$. In the previous section we showed the spectral function for $T_c/2$. As always we recover the AdS result of \eqqref{eq:spfAdS} in the UV limit. Furthermore, the quadratic dispersion in the long-wavelength limit, which follows from \eqqref{eq:largeNtpf}, is clearly visible here. These correspond to the Goldstone modes which arise from the breaking of the $U(N)$ symmetry to $U(N-1)$. The strongly overdamped sound-like mode, which is normally present in the intrinsic spectral function of a superconducting order parameter, is $1/N$ suppressed. This is because due to the large-$N$ limit, the spectral functions are only describing the fluctuations of the $N-1$ transverse order parameters and not the fluctuations of $O_1$. In addition, we observe that as the temperature decreases, the dissipation reduces. This follows from the quadratic dispersion becoming more narrow in the spectral function, as can be seen most clearly by comparing Fig. \ref{fig:sp2} with Fig. \ref{fig:sp152} .

\begin{figure}[!t]
\includegraphics[width=.95\linewidth,clip]{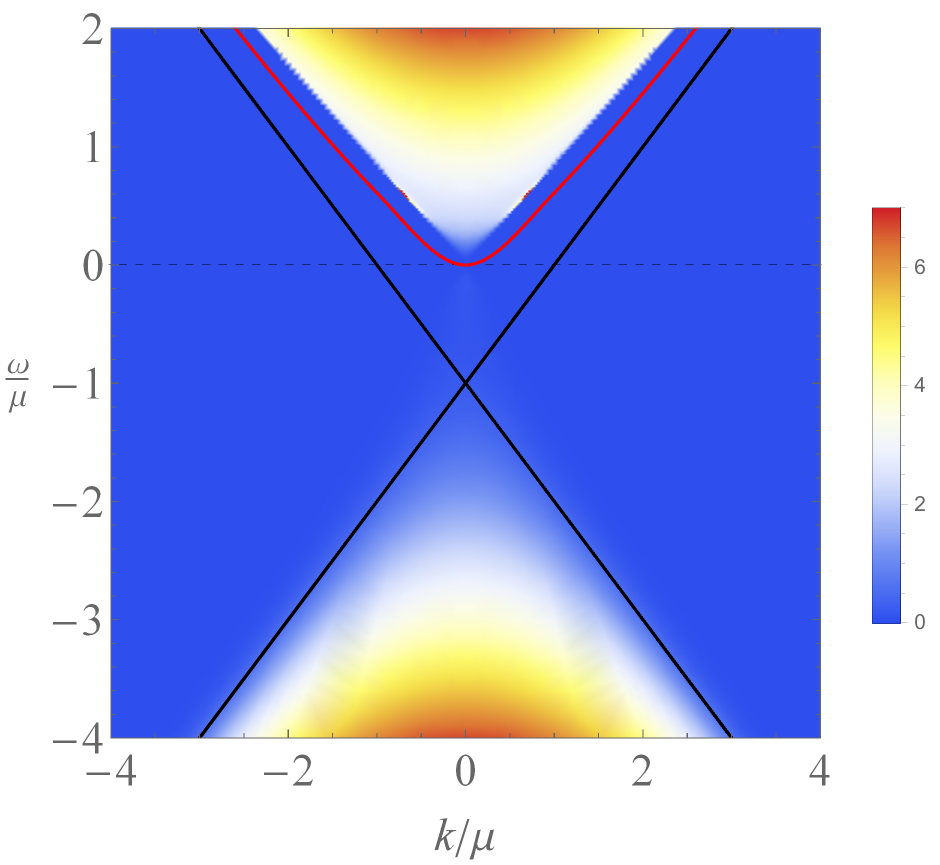}
\caption{\label{fig:sp0}The spectral function for $T=0.01T_c$, $q=3$, and $m^2=-3.5$. The quadratic dispersion is infinitely narrow at zero temperature, but was made visible by adding a small imaginary part to the frequency.}
\end{figure}

At $T=0$, we find numerically that the imaginary part of $a$ vanishes. This is to be expected, as the system becomes dissipationless at zero temperature. Consequently, in the $T=0$ intrinsic spectral function shown in Fig. \ref{fig:sp0} , the quadratic dispersion is a very long-lived mode. It is visible in the figure only because we added a small imaginary part to the frequency. Alternatively, we have checked that it is visible as a pole in the real part of the retarded Green's function, consistent with the Kramers-Kronig relations. Furthermore, we see that the linear second-sound mode which is normally present in the spectral function of the zero-temperature intrinsic order parameter dynamics not present.

Our numerics show that the real part of the coefficient $a$ does not vanish. Contrariwise, in BCS theory we typically expect a term proportional to $\omega^2$ rather than $\omega$ in the action at zero temperature \cite{UQF}. In the BEC regime a term linear in $\omega$ would be present. This suggests that we are describing a superfluid which resides outside the BCS regime. The value of the real part of $a$ can be related to the number density of the system. This is achieved by noting that the action should contain the topological term
\be{
S_{top}=-\int\dd t\dd \vec{x} \exv{n}\dot{\theta},}
where $\theta$ is the phase which is conjugate to the total number of particles $\int\dd \vec{x} \exv{n}$.  On the other hand, the dynamics of the two-point functions we obtain show that the action contains a term
\be{
-\int\dd t\int\dd^3\vec{x} O^*a (i\partial_t)O=a\int\dd t\int\dd^3\vec{x}\exv{O}^2\dot{\theta}+\dots,}
where we used $O=Ae^{i\theta}$ and expanded around the expectation value of $O$ using $\exv{A}=\exv{O}$ and $\exv{\theta}=0$. Comparing this term to the topological term above, we find that the coefficient $a$ should be given by
\be{ \label{eq:a}
a=-\frac{\exv{n}}{\exv{O}^2}.}
The right-hand side can be calculated from the bulk geometry without including fluctuations, whereas the left-hand side can be read off the Green's functions. We have performed this task for several values of $m^2$ and $q$ and checked that the result in \eqqref{eq:a} indeed holds.
\begin{figure}[!t]
\includegraphics[width=.95\linewidth,clip]{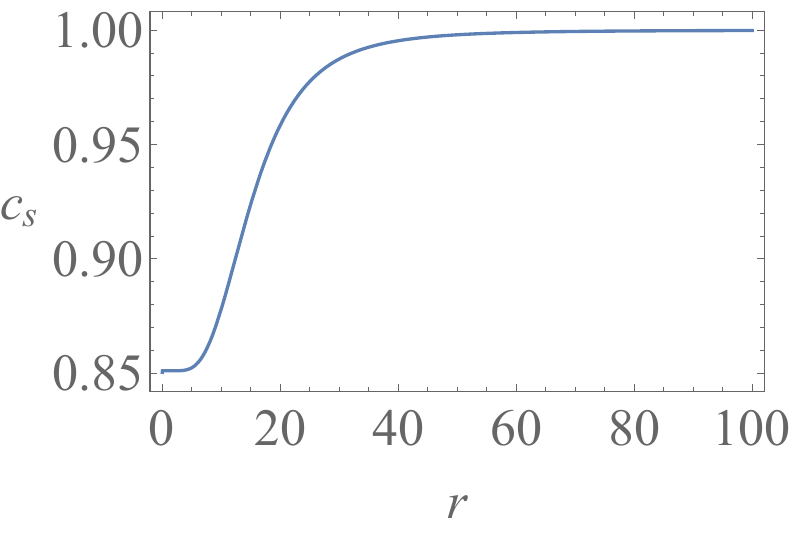}
\caption{\label{fig:c}The effective speed of light $c_s$ as a function of $r$. Here $q=3$ and $m^2=-3.5$.}
\end{figure}

Besides the quadratic dispersion, we see a cone appearing in the spectral functions. This is most clearly visible in the zero-temperature limit in Fig. \ref{fig:sp0} . Here, the spectral weight vanishes for $|\omega| <c_s |k| $ for a $k$-dependent $c_s$. This cone can be explained physically by the fact that the system contains fermionic species that are not gapped out by the nonzero expectation value $\exv{O_1}$. These species behave like free fermions with a Fermi velocity different from $c=1$ due to strong interactions. At large momenta, where we recover the AdS result, and the effective speed $c_s$ approaches 1, i.e., the speed of light. This is of course what we should expect from Lorentz invariance. For small momenta and frequency, we find that $c_s\approx 0.85$. Using the metric \textit{Ansatz} in \eqqref{eq:metricAnsatz}, we have calculated the effective speed of light in the bulk geometry as a function of $r$. This is given by
\be{
c(r)=\sqrt{\frac{f(r)e^{-\chi(r)}}{r^2}},}
where we use the zero-temperature numerical solution similar to Refs. \cite{ZeroT,Vegh}. As the result in Fig. \ref{fig:c} shows, the effective speed of light in the infrared part of the bulk geometry agrees with the value we obtained from the cone in Fig. \ref{fig:sp0} . Upon increasing the momentum, $c_s$ increases such that it asymptotically approaches the shifted light cone $|\omega + \mu|=|k|$, which indeed asymptotes to $c_s=1$ when $\omega$ and $k$ become large compared to $\mu$. Notice that for negative frequencies, the spectral weight inevitably becomes nonzero outside the shifted light cone. This is in principle possible because of the nonzero chemical potential. However, the spectral weight outside this cone is so small that it is hardly visible in Fig. \ref{fig:sp0} .
\begin{figure}[!t]
\includegraphics[width=.95\linewidth,clip]{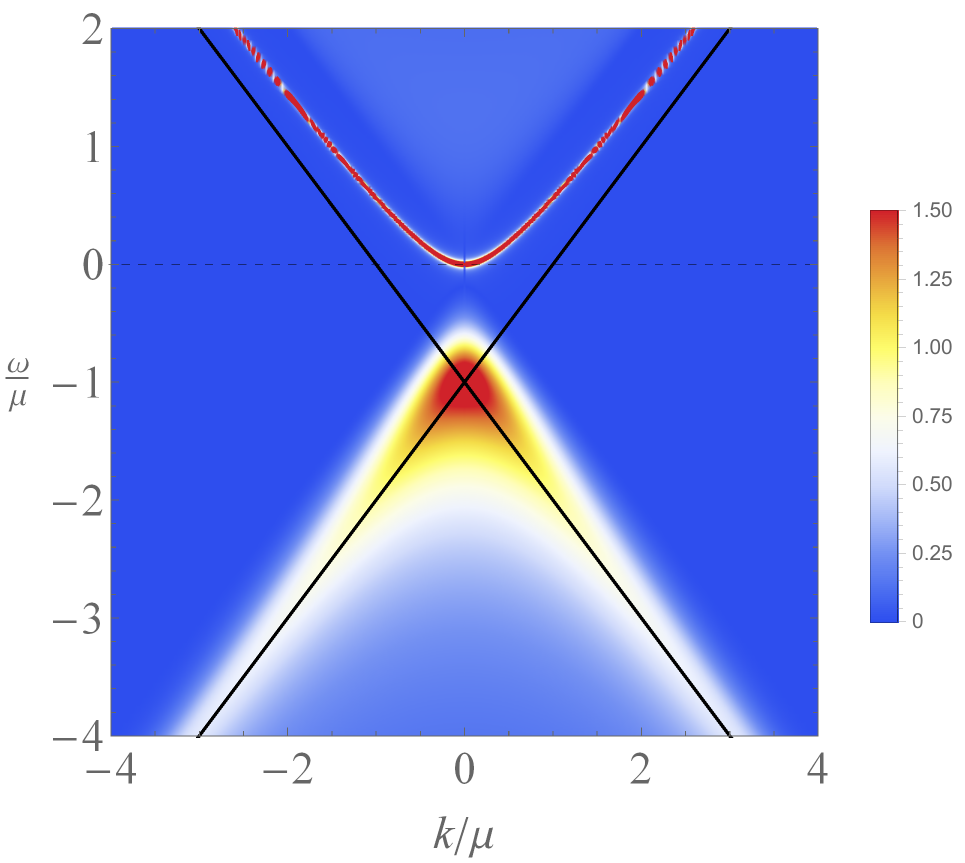}
\caption{\label{fig:AQ}The spectral function for $T=0$, $q=3$, and $m^2=-3.5$ using alternative quantization. We can clearly see the difference in the UV behavior. The quadratic dispersion is infinitely narrow, but was made visible by adding a positive imaginary part to the frequency.}
\end{figure}
Although the cone is most clearly visible at zero temperature, it is also present at higher temperatures. We can see it for example in Fig. \ref{fig:sp4} for $T=T_c/2$. However, the sharp transition to the region where the spectral function becomes nonzero, which is present for $T=0$, is smoothed out here by thermal fluctuations. The cone also appears in the alternative quantization scheme, as is shown in Fig. \ref{fig:AQ} . The appearance of the cone in the spectral densities is not predicted by the large-$N$ Ginzburg-Landau model. This model only includes the leading order terms in the long-wavelength and low-frequency expansion, which yield the Goldstone mode with a quadratic dispersion. We can indeed observe that this dispersion always lies outside the cone, meaning that it corresponds to smaller values of $\omega$ for a given $k$. There might be terms of higher order in the long-wavelength expansion which make the cone structure manifest. For example, consider a term of the form $(\sqrt{-\omega^2+c_s^2k^2})^\sigma$ in the denominator of the two-point function. Here $\sigma$ is a real number which should equal $-\nu$ in the AdS limit. As mentioned previously, the physical grounds for such a term are presumably due to the fact that the nonzero value of $\exv{O_1}$ does not gap out all the fermionic species in the dual field theory, similar to what happens in the two-flavor superconducting (2SC) phase of a quark-gluon plasma.

\section{Conclusions and Discussion} \label{sec:CD}
In this work we revisited the holographic superconductor and studied the intrinsic order parameter fluctuations, including backreaction in the bulk geometry. Our main results concern the intrinsic spectral functions below the critical temperature, for which we have presented a large-$N$ version of the Ginzburg-Landau model, which agrees with our numerical results in both the normal and superfluid phases. The low-temperature behavior of the spectral functions are reminiscent of a relativistic multi-component superfluid in the universal regime of the BEC-BCS crossover. In particular, due to the large-$N$ limit, the spectral functions only describe the transverse fluctuations of the multi-component order parameter. Consequently, only a Goldstone mode with a quadratic dispersion appears, and the Higgs mode and second-sound mode are absent.

A natural question which arises after this work is its relation and consistency with previous studies on this matter. In Ref. \cite{SchalmZaanen}, the order parameter fluctuations of the normal phase have been modeled to the standard Ginzburg-Landau action for $\textbf{k}=\textbf{0}$. We have seen that this is indeed possible also for nonzero $\vec{k}$ within the long-wavelength limit, since the large-$N$ Ginzburg-Landau model reduces to the standard Ginzburg-Landau model in the normal phase. Furthermore, Refs. \cite{Maeda} and \cite{Dector} study the poles of the static correlation function of the order parameter and find nontrivial poles in the complex $k$-plane. Since the static correlation function always has a pole at $k=0$ as well, this does not contradict the fact that we find no massive modes \cite{fn7}.
\begin{figure}[!t]
\subfloat[$g=0$\label{fig:quadmode}]
	{\includegraphics[trim = 0cm 0.1cm 0cm 0cm, clip=true, scale=.48]{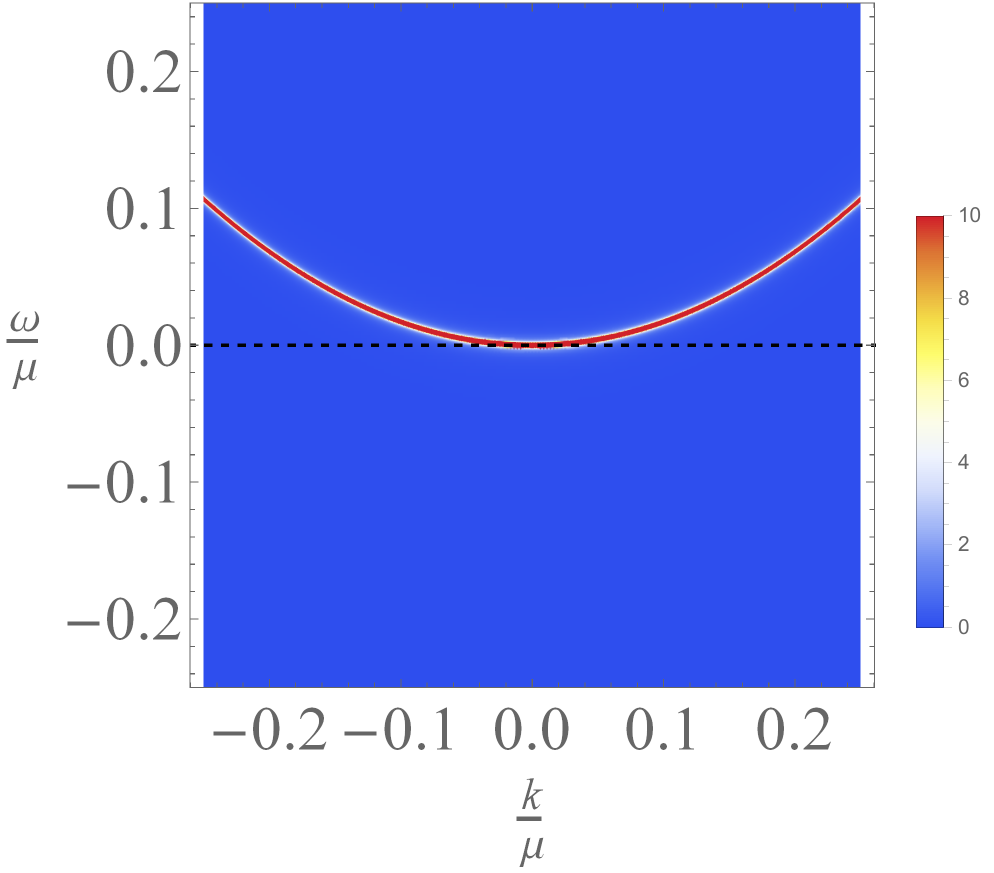}}

\subfloat[$g=1$\label{fig:linmode}]
	{\includegraphics[trim = 0cm 0.1cm 0cm 0cm, clip=true, scale=.48]{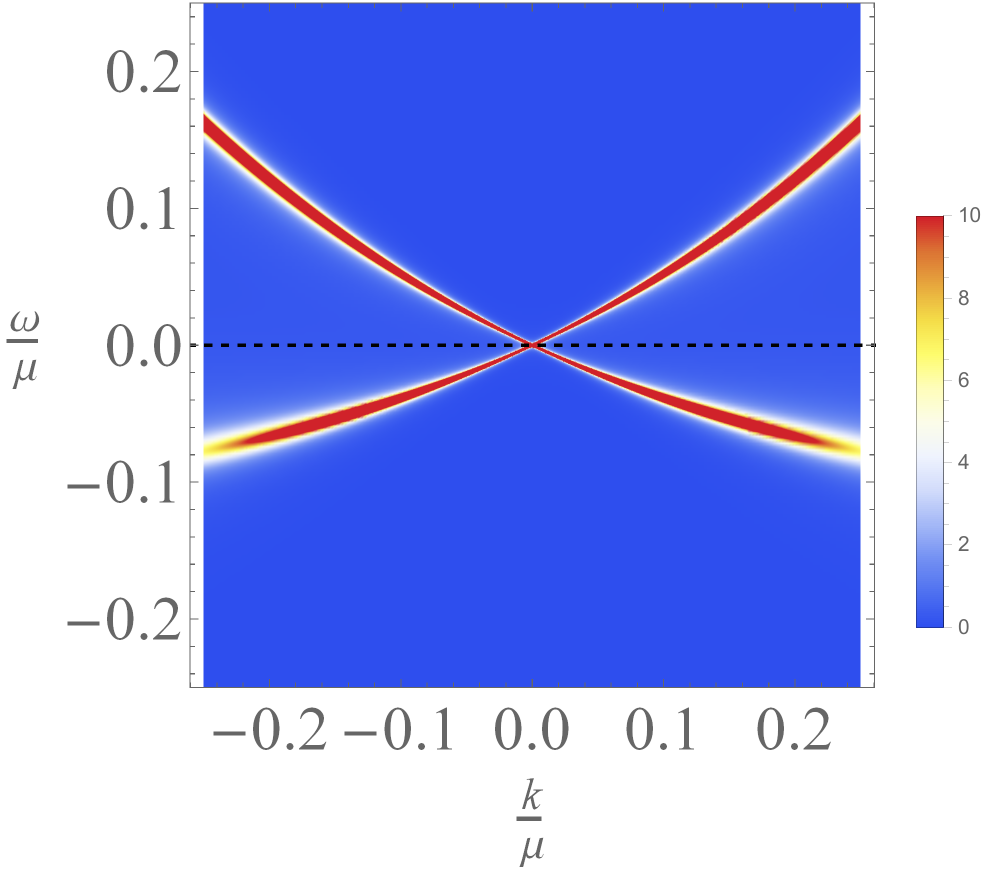}}
	\caption{The spectral function corresponding to the effective toy model in \eqqref{eq:toy}. Here we have taken $g=0$ in (a) and $g=1$ in (b), which shows the effect of integrating out the diffusive hydrodynamic mode. Furthermore, in both figures we have taken $D=1/4$ and the zero-temperature values of $a$ and $\gamma$ obtained from Fig. \ref{fig:sp0} .}
\label{fig:coupling}
\end{figure}
Finally, in Refs. \cite{Landsteiner,HerzogA} the quasinormal modes of the correlation functions are studied within the probe limit. As fluctuations of both the gauge field and the scalar field are taken into account there, a linear mode resembling the second-sound mode is found. Qualitatively, we indeed expect that including gauge field fluctuations leads to a linear mode, because of the coupling between the quadratic Goldstone mode we found above and the diffusive mode in the fluctuations of the current $J^\mu$ dual to the $U(1)$ gauge field. Phenomenologically, the linear mode can be obtained by realizing that these two modes are coupled by a current-current interaction of the form $q(O^*\partial_\mu O)J^\mu$, which leads to the following effective toy model for the order parameter correlation function after integrating out the $U(1)$ current fluctuations:
\be{ \label{eq:toy}
-i\exv{{O'}^*O'}^{-1}(\omega,\vec{k})=a\omega+\gamma|\vec{k}|^2+g^2\frac{-\omega^2+|\vec{k}|^2}{\omega+i D |\vec{k}|^2}.}
The first two terms on the right-hand side represent the Goldstone mode in the intrinsic order parameter dynamics. The last term gives the correction to the effective action after integrating out the diffusive hydrodynamic mode. Here we have introduced the coupling constant $g$ and the diffusion constant $D$. It follows from the form of the coupling as discussed above that $g$ is proportional to $q\exv{O}$, which yields the correct dependence of the speed of sound of the linear mode on the order parameter. Note that we did not include the dissipative cone structure in our toy model, because the terms in \eqqref{eq:toy} capture all the low-energy physics relevant in the argument we wish to make here.

The resulting spectral function is shown in Fig. \ref{fig:coupling} , where it is clearly visible that the quadratic mode now becomes linear in the long-wavelength limit. The corresponding speed of sound following from \eqqref{eq:toy} is
\be{
c_s=\text{Re}\lr{\frac{1}{\sqrt{1-a/g^2}}}}
which indeed vanishes for $g=0$ and is proportional to $q\exv{O}$ near the critical temperature. Nevertheless, we have argued that this is not the second-sound mode, since it does not correspond to the phase of the condensate but rather to the Goldstone mode of the transverse fluctuations of the multi-component order parameter. It would be interesting to see whether this sound-like mode, together with the first-sound mode, would indeed show up in a computation of the fully backreacted spectral functions. For this we would have to consider the coupled system of order parameter, $U(1)$ current and energy-momentum fluctuations, which can be obtained by including fluctuations of the gauge field and the metric. However, in order to derive a local time-dependent Ginzburg-Landau theory for the order parameter, as we have done in this work, we must consider only the intrinsic dynamics of the order parameter. Hence we do not couple to the other hydrodynamic degrees of freedom, because when integrating out these degrees of freedom the action becomes non-analytic in frequency and momentum space, as can be seen from the diffusive pole in \eqqref{eq:toy}. As a consequence, it would not be possible to use a gradient expansion for the action.

\appendix*

\section{Linearized equations of motion and retarded Green's function} \label{app}

Here, we demonstrate how to obtain the linearized equations of motion and the retarded Green's function from the action expanded up to second order in the fluctuations. As an example, we take the action considered in this paper within the probe limit. This corresponds to taking $q\rightarrow\infty$ while keeping $q\phi$ and $qA_\mu$ constant. The metric then decouples from the scalar field and the gauge field. Although this should only be a good approximation sufficiently close to the critical temperature, where the backreaction of the scalar field is negligible, this only serves as an example of how to calculate the full retarded Green's function. It should be clear (but tedious) how to extend this to include the metric fluctuations.

Since we work in the probe limit, the relevant action to consider is given by
\ba{ \nonumber
S=&\int\dd^{5}x\sqrt{-g}\lr{-\frac{1}{4}F^2-|D\phi|^2-m^2|\phi|^2}\\ &-\int\dd^4 x\sqrt{-h}\Delta_- |\phi|^2.\label{eq:Sprobe}}
The last term is evaluated at the boundary $r=\infty$, where $h$ is the determinant of the induced metric. This term serves as a counterterm that renders the on-shell action finite. Moreover, it is compatible with the identification of $\phi_s$ in \eqqref{eq:phiexp} as a source. Different boundary terms are needed when working in alternative quantization or in the canonical ensemble, where we fix the charge density rather than the chemical potential on the boundary \cite{HHH2}.

We proceed by expanding each term up to second order in the fluctuations of the fields. Using Stokes theorem, the second and third term in \eqqref{eq:Sprobe} yield the second-order terms 
\ban{
S^{(2)}_{KG}&=-\int \dd^5 x \sqrt{-g} \Big\{D_\mu\delta\phi(D^\mu\delta\phi)^*+m^2\delta\phi^*\delta\phi\\ &\qquad\qquad\qquad\quad -q^2\delta A_\mu\delta A^\mu|\phi|^2 \\
&\qquad\qquad\qquad\quad+iq\delta A_\mu\big[\delta\phi (D^\mu\phi)^*+\phi(D^\mu\delta\phi)^*\\&\qquad\qquad\qquad\qquad\qquad\quad-\phi^* D^\mu\delta\phi-\delta\phi^* D^\mu\phi\big]\Big\} \\
&=\int \dd^5 x \sqrt{-g} \bm{ \bigg(}\frac{1}{2}\delta\phi^*D^2\delta\phi+\text{h.c.}+m^2\delta\phi^*\delta\phi \\ &\qquad\qquad\qquad -q^2\delta A_\mu\delta A^\mu|\phi|^2 \\
&\qquad\qquad\quad-\big\{iq\delta A_\mu[(D^\mu\phi)^*-\phi^*D^\mu]\delta\phi+\text{h.c.}\big\}\bm{ \bigg)} \\
&\quad-\frac{1}{2}\int\dd^4 x\sqrt{-h}n^\mu\lr{\delta\phi^* D_\mu\delta\phi+\text{h.c.}} \\
&=\frac{1}{2}\int \dd^5 x \sqrt{-g} \bm{ \bigg(}\delta\phi^*D^2\delta\phi+\text{h.c.}+m^2\delta\phi^*\delta\phi\\
&\qquad\qquad\qquad\quad\,\,\, -2q^2\delta A_\mu\delta A^\mu|\phi|^2 \\
&\qquad\qquad\qquad\,\,\,-\big\{iq\delta \phi[2(D^\mu\phi)^*+\phi^*\nabla^\mu]\delta A_\mu\\
&\qquad\qquad\qquad\,\,\,+iq\delta A_\mu[(D^\mu\phi)^*-\phi^*D^\mu]\delta\phi+\text{h.c.}\big\}\bm{ \bigg)}  \\
&\quad-\frac{1}{2}\int\dd^4 x\sqrt{-h}n^\mu\lr{\delta\phi^* D_\mu\delta\phi-iq \delta A_\mu \phi^*\delta\phi+\text{h.c.}}.}
Here we introduced the unit normal $n^\mu=r\delta^\mu_r$. The contributions of the last term is
\ban{S^{(2)}_{\partial}&=-\int \dd^4 x \sqrt{-h} \Delta_-\delta\phi^*\delta\phi.}
Finally, the first term in \eqqref{eq:Sprobe} yields 
\ban{
S^{(2)}_{M}=&-\frac{1}{4}\int\dd^{5}x\sqrt{-g}\delta F_{\mu\nu}\delta F^{\mu\nu} \\
=&-\frac{1}{2}\int\dd^{5}x\sqrt{-g}\nabla_\mu\delta A_\nu \lr{\nabla^\mu\delta A^\nu-\nabla^\nu\delta A^\mu} \\
=&\frac{1}{2}\int\dd^{5}x\sqrt{-g}\delta A_\nu \lr{g^{\mu\nu}\square-\nabla^\mu\nabla^\nu}\delta A_\mu \\&-\frac{1}{2}\int d^4 x \sqrt{-h}\delta A_\nu\lr{g^{\mu\nu} n^\rho\nabla_\rho-n^\mu\nabla^\nu}\delta A_\mu.}
Here $\square\equiv \nabla_\mu\nabla^\mu$. Adding these contributions, we can write the second-order terms of the action as the sum of bulk and boundary terms: $S^{(2)}=S^{(2)}_{bulk}+S^{(2)}_{bdy}$. Here,
\ba{S^{(2)}_{bulk}=&\frac{1}{2}\int\dd^5 x\sqrt{-g}\nonumber \\
&\times\bm{ \bigg(}\delta\phi^*\lr{D^2-m^2}\delta\phi+\text{h.c.} \nonumber \\
&\qquad-\big\{iq\delta A_\mu[(D^\mu\phi)^*-\phi^*D^\mu]\delta\phi+\text{h.c.}\big\} \nonumber \\
&\qquad-\big\{iq\delta \phi[2(D^\mu\phi)^*+\phi^*\nabla^\mu]\delta A_\mu+\text{h.c.}\big\} \nonumber \\
&\qquad+\delta A_\nu[g^{\mu\nu}\lr{\square-2q^2|\phi|^2}-\nabla^\mu\nabla^\nu]\delta A_\mu\bigg),}
where $D^2\equiv D_\mu D^\mu$. We can simplify the above using the radial gauge $A_r=\delta A_r=0$. Moreover, we note that up to zeroth order, $\phi^*=\phi=\phi(r)$, $A=A_t(r)\dd t$, and the metric is given by the \textit{Ansatz} in \eqqref{eq:metricAnsatz}. This yields 
\ba{
S^{(2)}_{bulk}=&\frac{1}{2}\int\dd^5 x\sqrt{-g}\nonumber\\
&\times\bigg\{\delta\phi^*\lr{D^2-m^2}\delta\phi+\text{h.c.} \nonumber \\
&\qquad+[\delta A_a\lr{2q^2A^a\phi+iq\phi \partial^a}\delta\phi+\text{h.c.}] \nonumber \\
&\qquad+[\delta \phi\lr{2q^2A^a\phi-iq\phi\partial^a}\delta A_a+\text{h.c.}] \nonumber \\
&\qquad+\delta A_a\big[g^{a\nu}\lr{\square-2q^2\phi^2}-\nabla^{\nu}\nabla^a\big]\delta A_{\nu}\bigg\},}
where the Latin indices run over the coordinates $t$ and $\vec{x}$. From $S^{(2)}_{bulk}$, we can then read off the linearized equations of motion. Writing these in the form \eqqref{eq:leom} yields 
\begin{widetext}
\be{
\label{eq:leom2}-
\begin{pmatrix} D^2-m^2 & 0 & 2q^2A^{\nu}\phi+iq\phi\partial^{\nu}\\
0 & D^{*2}-m^2 & 2q^2A^{\nu}\phi-iq\phi\partial^{\nu}\\
2q^2A^a\phi+iq\phi\partial^a & 2q^2A^a\phi-iq\phi\partial^a& g^{a\nu}\lr{\square-2q^2|\phi|^2}-\nabla^{\nu}\nabla^a\\
\end{pmatrix}\begin{pmatrix}\delta \phi \\ \delta \phi^* \\ \delta A_{\nu} \end{pmatrix}=\vec{0},}
\end{widetext}
from which we can read off the matrix operator $\textbf{G}_B^{-1}$ that was introduced in \eqqref{eq:S2}. Here we defined $D^*_\mu\equiv\nabla_\mu+iq A_\mu$. Writing the field fluctuations in Fourier space, this indeed yields a coupled set of ordinary differential equations. 

From the off-diagonal elements in \eqqref{eq:leom2}, it follows that the equations decouple in the normal phase, since then $\phi=0$ to zeroth order. The intrinsic dynamics are then the same as the full dynamics. Moreover, it follows that we can view the parameter $q\phi$ as an expansion parameter, in the sense that the intrinsic dynamics are a good approximation for the full dynamics for small values of this parameter $q\phi$. This corresponds to temperatures near the critical temperature. However, for the full spectral functions of the theory, we in general cannot ignore the gauge field fluctuations. Indeed, the off-diagonal terms have important effects especially in the long-wavelength limit, as can be seen in the toy model presented in Section \ref{sec:CD} in Fig. \ref{fig:sp0}. Nevertheless, as we argued in this paper, the intrinsic dynamics of the order parameter already allows us to extract important information about the nature of the order parameter.

The boundary term is given by
\ba{
S^{(2)}_{bdy}=-\frac{1}{2}\int\dd^4x \sqrt{-h}&\bigg[\delta\phi^*\lr{n^\mu D_\mu +\Delta_-}\delta\phi+\text{h.c.} \nonumber \\
&-iqn^\mu\delta A_\mu\phi^*\delta\phi+\text{h.c.} \nonumber \\
&+\delta A_\nu\lr{g^{\mu\nu} n^\rho\nabla_\rho-n^\mu\nabla^\nu}\delta A_\mu\bigg].}
From this we can read off the expressions for the retarded Green's functions. Using the radial gauge and the fact that $n^\mu=r\delta^\mu_r$, we have that $n^\mu D_\mu=r\partial_r$ and $n^\mu\delta A_\mu=0$. Inserting the expansions from \eqqref{eq:Phiexp} and neglecting terms which vanish as $r\rightarrow\infty$, we obtain that
\ba{
S^{(2)}_{bdy}=\frac{1}{2}\frac{1}{(2\pi)^4}\int\dd^4k \bigg[&2\nu\lr{\delta\phi_s^*\delta\phi_v+\text{h.c.}}\nonumber \\ &+2\eta^{ab}\delta A_{a(s)}\delta A_{b(v)}\bigg].}
Writing this in the same form as in \eqqref{eq:S2bdy}, we read off the retarded Green's function
\be{ \label{eq:appGR}
\textbf{G}_R=\begin{pmatrix} 2\nu \frac{\partial\delta\phi_v}{\partial\delta\phi_s} & 2\nu \frac{\partial\delta\phi_v}{\partial\delta\phi_s^*} & 2\nu \frac{\partial\delta\phi_v}{\partial\delta A_{b(s)}}\\
2\nu \frac{\partial\delta\phi_v^*}{\partial\delta\phi_s} & 2\nu \frac{\partial\delta\phi_v^*}{\partial\delta\phi_s^*} & 2\nu \frac{\partial\delta\phi_v^*}{\partial\delta A_{b(s)}}\\
2 \eta^{ac} \frac{\partial\delta A_{c(v)}}{\partial\delta\phi_s} & 2 \eta^{ac}\frac{\partial\delta A_{c(v)}}{\partial\delta\phi_s^*} & 2 \eta^{ac}\frac{\partial\delta A_{c(v)}}{\partial\delta A_{b(s)}}\\
\end{pmatrix}.}
When taking these derivatives, all other sources are kept constant, as explained in Subsection \ref{ss:FD} . Thus, each column can be found from a solution which sources only one of the field fluctuations on the boundary. The result is as expected, e.g., to obtain the two-point function $\exv{{O'}^*O'}$ we take the variation of the vacuum expectation value $\exv{O}$ to the source $\phi_s$, as in \eqqref{eq:varphi}. This is done by using a particular solution to the linearized equations of motion in \eqqref{eq:leom2}, as also explained in Subsection \ref{ss:FD} .

Notice that the coupling of the operators in the field theory is due to the coupled linearized equations of motion in \eqqref{eq:leom2} in the bulk. If we wish to compute the intrinsic dynamics of the order parameter, rather than solving \eqqref{eq:leom2}, we solve \eqqref{eq:probeEOM}. Notice that in this case this also means that $\delta\phi$ and $\delta\phi^*$ are decoupled. This is shown in the matrix in \eqqref{eq:leom2}, where $\delta\phi$ and $\delta\phi^*$  are coupled not directly but only indirectly through $\delta A_a$, so that this coupling vanishes when considering the intrinsic order parameter fluctuations.

\acknowledgments

It's a pleasure to thank the participants of the Chalmers workshop "Applications of the Gauge/Gravity Duality 2015", where this work was first presented, for discussions and feedback. Moreover, we would like to thank Christopher Herzog and Daniel Arean for their helpful comments. This work was supported by the Stichting voor Fundamenteel Onderzoek der Materie (FOM) and is part of the D-ITP consortium, a program of the Netherlands Organisation for Scientific Research (NWO) that is funded by the Dutch Ministry of Education, Culture and Science (OCW).


\begin{thebibliography}{99}
\bibitem{fn1}
Although these fermion gases at unitarity consist of Cooper pairs in a condensed state, the Cooper pairs consist of neutral fermions. Therefore the term superfluid may be more appropriate.
\bibitem{fn2}
In alternative quantization, $\phi_v=0$ is used instead. This is only possible when the term with $\phi_s$ in \eqqref{eq:phiexp} has a normalizable fall-off.
\bibitem{fn5}
The frequency and the wavevector are made dimensionless in the way consistent with \eqqref{eq:dimlessunits}, i.e., $\tilde{\omega}=\omega L/c$ and $\tilde{\vec{k}}=\vec{k}L$.
\bibitem{fn6}
With this particular solution, we can also find other correlators that follow from expressions like $\frac{\delta A_{\mu,v}}{\delta\phi_s}$. In fact, with this solution we can find exactly one column of the retarded Green's function matrix.
\bibitem{fn7}
E.g., the function $(\xi^2k^4+k^2)^{-1}$ has a pole at $k=\pm i/\xi$ but does not describe a massive mode.
\end{thebibliography}
\end{document}